\documentclass[10pt]{article}

\usepackage[margin=1in]{geometry}  
\usepackage[T1]{fontenc}          
\usepackage{algorithm}
\usepackage{algpseudocode}
\usepackage{amsmath}
\usepackage{graphicx}
\usepackage{times}                
\usepackage[export]{adjustbox}
\usepackage{float}
\usepackage{xr}
\makeatletter
\newcommand*{\addFileDependency}[1]{
\typeout{(#1)}
\@addtofilelist{#1}
\IfFileExists{#1}{}{\typeout{No file #1.}}
}\makeatother

\newcommand*{\myexternaldocument}[1]{%
\externaldocument{#1}%
\addFileDependency{#1.tex}%
\addFileDependency{#1.aux}%
}

\myexternaldocument{supporting_material}

\title{Dynamic-Computed Tomography Angiography for Cerebral Vessel Templates and Segmentation}
\author{
  Shrikanth Yadav$^{1,2}$\\
  Jisoo Kim, MD$^{2,3}$\\
  Geoffrey Young, MD$^{2,3}$\\
  Lei Qin, PhD$^{1,3}$\\[8pt]
  \small
  $^{1}$Dana-Farber Cancer Institute, Boston, MA \\
  $^{2}$Brigham and Women's Hospital, Boston, MA \\
  $^{3}$Harvard Medical School, Boston, MA
}
\date{} 

\begin{document}
\begin{titlepage}

\begin{center}
    {\large \textbf{Dynamic-Computed Tomography Angiography for Cerebral Vessel Templates and Segmentation}} \\[2em]
    \normalsize
    Shrikanth Yadav$^{1,2}$, 
    Jisoo Kim, MD$^{2,3}$, 
    Geoffrey Young, MD$^{2,3}$,
    Lei Qin, PhD$^{1,3}$ \\[2em]

    \small
  $^{1}$Dana-Farber Cancer Institute, Boston, MA \\
  $^{2}$Brigham and Women's Hospital, Boston, MA \\
  $^{3}$Harvard Medical School, Boston, MA
\end{center}

\noindent
\textbf{Corresponding Author:}\\
Lei Qin, PhD\\
Email: \texttt{lei\_qin@dfci.harvard.edu}
\end{titlepage}

\section*{Abstract}

\noindent \textbf{Background}: Computed Tomography Angiography (CTA) is crucial for cerebrovascular disease diagnosis. Dynamic CTA is a type of imaging that captures temporal information about the We aim to develop and evaluate two segmentation techniques to segment vessels directly on CTA images: (1) creating and registering population-averaged vessel atlases and (2) using deep learning (DL). 

\noindent \textbf{Methods}: We retrieved 4D-CT of the head from our institutional research database, with bone and   soft   tissue subtracted from post-contrast images.  An Advanced Normalization Tools pipeline was used to create angiographic atlases from 25 patients. Then, atlas-driven ROIs were identified by a CT attenuation threshold  to generate segmentation of the arteries and veins using non-linear registration. To create DL vessel segmentations, arterial and venous structures were segmented using the MRA vessel segmentation tool, iCafe, in 29 patients. These were then used to train a DL model, with bone-in CT images as input.  Multiple phase images in the 4D-CT were used to increase the training and validation dataset. Both segmentation approaches were evaluated on a test 4D-CT dataset of 11 patients which were also processed by iCafe and validated by a neuroradiologist. Specifically, branch-wise segmentation accuracy was quantified with 20 labels for arteries and one for veins.

\noindent \textbf{Results}: DL outperformed the atlas-based segmentation models for arteries (average modified dice coefficient (amDC) 0.856 vs. 0.324) and veins (amDC 0.743 vs. 0.495) overall. For ICAs, vertebral and basilar arteries, DL and atlas -based segmentation had an amDC of 0.913 and 0.402, respectively. The amDC for MCA-M1, PCA-P1, and ACA-A1 segments were 0.932 and 0.474, respectively. 

\noindent \textbf{Conclusions}: Angiographic CT templates are developed for the first time in literature. However, poor registration-based vessel segmentation performance highlights the challenge of anatomical variability across individuals. Using 4D-CTA enables the use of tools like iCafe, lessening the burden of manual annotation.

\section{Introduction}

Intracranial computed tomographic angiography (CTA), employing intravenous contrast agents to enhance the image contrast of blood vessels, is the first-line diagnostic technique for detection of abnormalities such as stenoses and aneurysms\cite{walluscheck_mr-ct_2023}. Since its introduction, several CTA techniques have been developed, including multiphase CTA. In multiphase CTA\cite{dundamadappa_multiphase_2021}, arterial, mid-venous, and late venous phase images are acquired. In dynamic 4D-CTA\cite{rajiah_dynamic_2022}, a continuous or intermittent CT acquisition of the head is acquired throughout the passage of contrast, yielding multiple images at different arterial and venous phases. As intracranial CTA is critical for diagnosis of emergent life-threatening conditions such as strokes and aneurysms, its interpretation requires substantial expertise and time. Consequently, development of robust image analysis techniques to assist expert radiologists promises great clinical impact. These efforts include automatic segmentation of the Circle of Willis\cite{yang_topcow_2024} to assess the angioarchitectures, and automated labeling of arterial vessels in CTA for patients with Stroke\cite{rist_bifurcation_2022} once vessel segmentations are available. 

Consequently, we leverage the fast acquisition times of 4D-CTA to address three challenges in the context of CT image analysis:
\begin{enumerate}
    \item \textbf{Vessel Templates:} In medical imaging, an anatomical template is the first step for understanding the normal morphology\cite{nowinski_evolution_2021}. For magnetic resonance angiograhphy (MRA), a brain artery template was developed\cite{mouches_statistical_2019}, showcasing the normal morphology of arterial structures. The development of similar atlases and tools for vessel segmentation in CTA could significantly improve clinical diagnosis and research, particularly in the emergency department and acute-care settings, because it is far more widely available, faster, and more accessible than MRA\cite{noauthor_comparing_nodate}. However, unlike MRA, which produces near-perfect background suppression and vessel segmentation during image acquisition, CTA images require substantial post-processing to remove background bone and soft tissue and isolate arteries from veins.  To date, this remains a time-consuming process requiring human expertise. No CT vessel atlas has been reported in the literature, and only two non-angiographic CT brain atlases available. The first template\cite{rorden_age-specific_2012} is derived from a cohort of 30 individuals in the Montreal Neurological Institute (MNI) space. The second \cite{muschelli_publicly_2020} is derived from a population of 130 individuals from the publicly accessible CQ500\cite{chilamkurthy_deep_2018} dataset. Both templates are derived from CT images that include bones and soft tissue. The major obstacle to creating angiographic atlases in CT is the overlapping image intensities between contrast-enhanced vessels and the background. Also, even with enhanced vessel contrast, template construction, and artery-vein separation are complicated by the intricate anatomy of the cerebral vasculature.  For instance, a portion of the internal carotid arteries (ICA) is surrounded by the cavernous sinuses (CVS), a venous structure. For clinical interpretation, CTA images are ideally acquired during the peak arterial phase before contrast opacification of the CVS and other veins. Still, in practice, the scan timing is often suboptimal, resulting in CTA images with varying degrees of vein `contamination'. 
    
    To produce distinct angiographic CT atlases for intracranial arteries and veins (to our knowledge, the first in the literature), we exploit the near-perfect background bone and soft-tissue subtracted CTA images from 4D-CTA.  We then apply a conditional subtraction approach to separate arteries and veins into distinct volumes and then apply a publicly available multivariate template construction pipeline to construct the angiographic templates. The utility of the template is shown using a registration-based segmentation. Specifically, we used non-linear registration to align the templates to new CTA images from a test Dynamic CTA cohort, demonstrating satisfactory segmentation of the ICA and the first-order branches of the middle, posterior, and anterior cerebral arteries.

    \item \textbf{Vessel Segmentation:} Given the wide clinical use of CTA imaging, there is an active effort to develop image processing techniques for CTA, particularly for automatic vessel segmentation\cite{yang_topcow_2024} and centerline extraction\cite{liu_automated_2024}. Automatic vessel segmentation is crucial for aiding radiologists in visualizing the vasculature, serving as a foundation for mathematical models of the cerebral vasculature\cite{talou_adaptive_2021}, and enhancing deep learning models to focus on vascular pathologies\cite{di_noto_towards_2023}. However, despite the significant advances in CTA, automatic and semi-automatic vessel segmentation tools for CTA lag behind those developed for Magnetic Resonance Angiography (MRA). This disparity is largely due to the superior contrast-to-background ratio in MRA compared to CTA. Consequently, MRA image analysis has benefited from the development of numerous datasets, such as the MRA annotations from the OASIS-3 dataset\cite{noauthor_oasis-3_nodate}, and sophisticated tools like iCafe\cite{chen_development_2018}. Similar comprehensive vessel segmentation tools for CTA have not been developed, highlighting a gap in CTA vascular image analysis.

    This work aims to develop and quantify automatic vessel segmentation in conventional CTA images by leveraging the properties of Dynamic CTA. Using preliminary vessel traces from iCafe, we train two deep learning models based on the nnU-net\cite{isensee_nnu-net_2021} and the NexToU\cite{shi_nextou_2023} frameworks. For quantify the performance of these models, creating accurate ground truth segmentations of arteries and veins presents several challenges, as outlined in a recent review\cite{zhao_segmentation_2019}. We address three major hurdles pertaining to the anatomical complexity, limited training data, and the need for arterial-venous separation: 

\begin{itemize}
    \item \textit{Anatomical complexity}: Expertise in anatomy is essential for annotating CTA images due to the presence of bone, soft tissue, and non-vascular contrast-dependent structures. We address this challenge by utilizing soft tissue subtracted images derived from dynamic CTA. The high temporal resolution of dynamic CTA ensures minimal motion artifacts in the subtracted images across different phases, thereby improving the contrast-to-background ratio.

    \item \textit{Training Data}: Deep learning models need training examples where ground truth segmentations are available. The software CerebralDoc\cite{fu_rapid_2020} used 18,766 CTA images that required expert annotation, which is a time-consuming and expensive process. More recently, virtual reality-based tools were used to create voxel-level annotations in the TopCoW challenge\cite{yang_topcow_2024}. However, this challenge was limited to anatomical regions around the circle-of-Willis (CoW) and did not extend to peripheral vessels. We address the problem of creating ground truth annotations using the semi-automatic vessel annotation tool, iCafe. Taking a step further, with our method using dynamic CTA, we leverage the heterogeneity of our dataset in terms of contrast phase and artifacts to create a more rigorous deep-learning vessel segmentation model. Consequently, during testing, our segmentation models were also evaluated for their ability to segment vessels with varying levels of voxel intensities. 

    \textit{Arterial-venous Separation}: Even with clinician experts, differentiating the arteries, veins, cavernous sinuses, and the skull base is challenging due to their proximity and caliber. Consequently, only modalities with high temporal resolution like dynamic CTA\cite{meijs_artery_2018} and digital subtraction angiography\cite{su_cave_2023} attempt artery-vein separation by exploiting temporal information. Using a simple conditional subtraction approach, we use time-resolved data to isolate arterial and venous structures into two different volumes.

\end{itemize}

\item \textbf{Extract Ground Truth Segmentations}: Extracting branch-specific ground truth segmentations of the vessel in CTA is a challenging and expensive task given the necessary time and expertise necessary, respectively \cite{fu_rapid_2020}. However, bone and soft tissue subtracted images from the input Dynamic 4D-CTA dataset enabled cross-modality application of the MRA vessel analysis tool iCafe\cite{chen_development_2018} to our CTA cohort. Specifically, iCafe was developed for performing semi-automatic annotations and feature extraction of arterial branches from MRA images. We use the subtracted series from our test Dynamic CTA cohort to create ground truth (GT) segmentations of the arteries and veins. These annotations were exported into 3D segmentations and validated by expert radiologist co-investigators. These segmentations were then used to evaluate the accuracy of the registration-based segmentation using our template, and the deep learning model.

\end{enumerate}

Consequently, the manuscript is organized as follows. Section~\ref{sec:methods} describes the acquisition and preprocessing of 
data, explains the multivariate template creation, and outlines how we train our deep learning models. We then detail our approach for evaluating the resulting vessel segmentations, including artery and venous ground truth creation, template-based region-of-interest (ROI) validation, and deep learning model validation. In Section~\ref{sec:results}, we present the resulting vessel templates, followed by quantitative assessments of the registration-based segmentation and the performance of our deep learning models. Section~\ref{sec:discussion} interprets these findings, focusing on both the vessel template construction and the deep 
learning–based segmentation. Finally, we conclude with Section~\ref{sec:conclusion}, summarizing the main contributions and 
discussing directions for future work.

\section{Methods }
\label{sec:methods}

\subsection{Data}

\begin{figure} 
   \begin{center}
   \includegraphics[clip, trim=0in 1.2in 0in 1.25in,width=\textwidth]{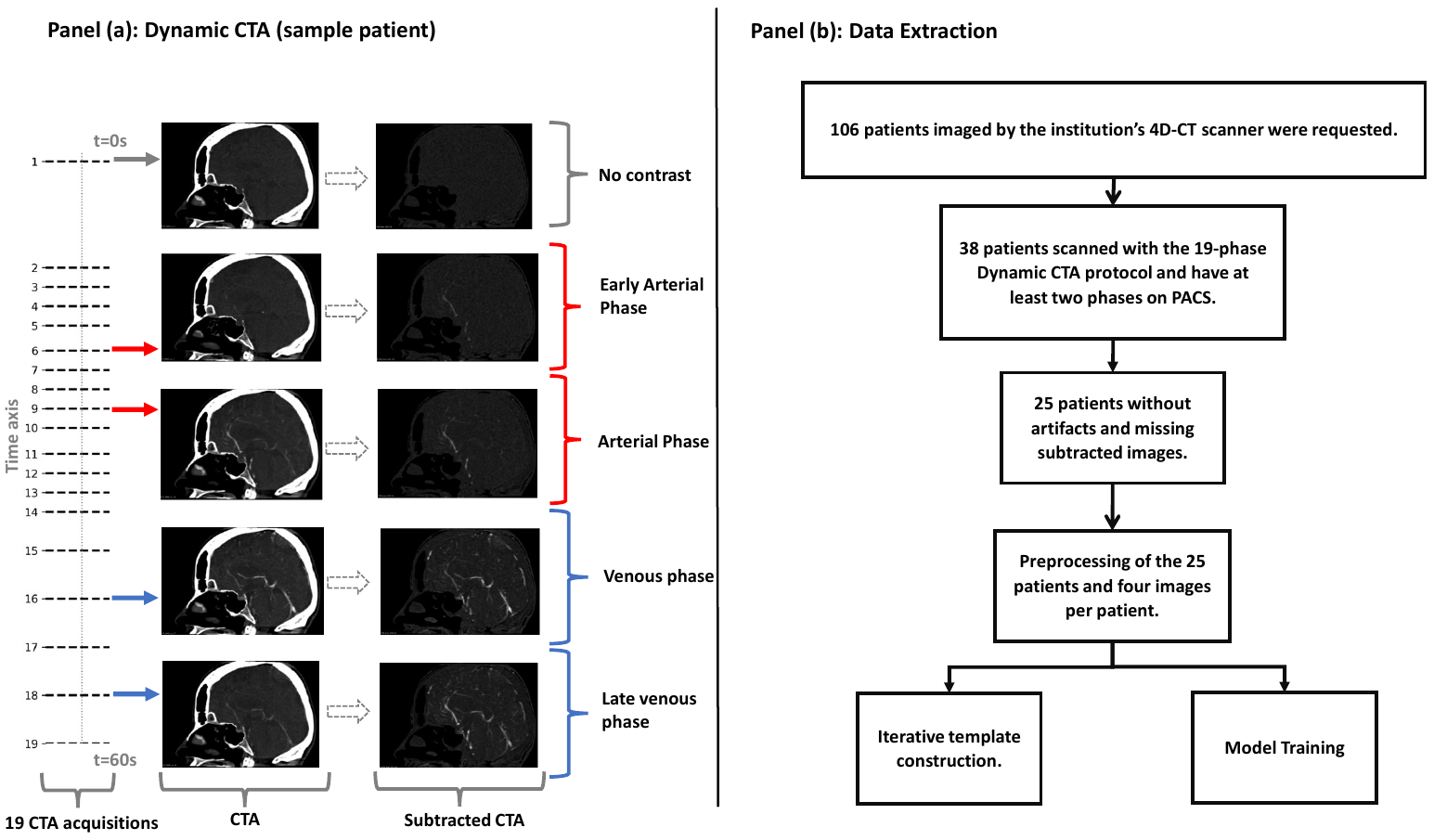}
   \caption{Panel (a) Dynamic CTA from a sample patient. The axis on the left shows the acquisition time of each phase; the red and blue arrows point to the arterial phase and venous phase images respectively. Panel (b) provides an overview of the cohort selection process for constructing the template and training the deep learning model. 
   \label{fig:patient_selection} 
    }  
    \end{center}
\end{figure}

\begin{figure} 
   \begin{center}
   \includegraphics[clip, trim=0in 4.8in 0in 2.5in,width=\textwidth]{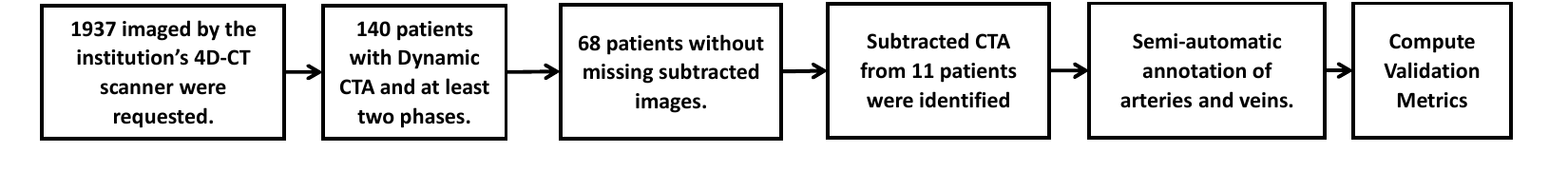}
   \caption{Flowchart illustrating the evaluation data selection.From an initial pool of 1937 scans performed on the institution’s 4D-CT scanner, we identified 140 patients with Dynamic CTA (having at least two phases). After excluding those lacking subtracted images, 68 patients remained; of these, 11 provided subtracted CTA suitable for semi-automatic artery/vein annotation. \label{fig:patient_selection_eval} 
    }  
    \end{center}
\end{figure}

The 4D-CTA, retrieved from our clinical research data warehouse, was performed on a 320-detector-row CT scanner (Canon-Toshiba Aquilion One. Toshiba Medical Systems, Tokyo, Japan). For each study, image acquisition started roughly 7 seconds after the start of a 4-5mL/s, 75-100 mL, bolus injection of iodinated contrast (Omnipaque 350 General Electric Healthcare; Chicago, USA) and continued intermittently throughout baseline, arterial and venous phases. Shown in the left column of Figure~\ref{fig:patient_selection} Panel (a), images acquired at 19 time-points over 60 seconds.
The scanner uses the first image, acquired before the arrival of contrast in the brain, as the baseline for the subtraction of bone and soft tissue from the later phase images (right column in Figure~\ref{fig:patient_selection}(a)). In this study, we note that the peak arterial and venous phase images are of primary interest. In Figure~\ref{fig:patient_selection}(a), this corresponds to $t=27.6$s and $t=45.3$s respectively. Scan parameters were: axial mode, 80kV,  150mA,  0.75s rotation time, 320 slices with 0.5 mm thickness, and matrix 512 $\times$ 512.

Panel (b) of Figure~\ref{fig:patient_selection} depicts the patient selection processes used for constructing template and the training data. 106 patients scanned on the Toshiba Aquillion One scanner in 2023 were retrieved from PACS. Of these, 38 patients had at least two phases of scans with acceptable image quality. After reviewing arterial and venous phase images to exclude those with artifacts, 25 patients were chosen to construct the template. 

Figure~\ref{fig:patient_selection_eval} shows the patient selection criteria from the evaluation dataset. Here, 1910 patients scanned by the Toshiba Aquillion One scanner from 2016 to 2022 were initially retrieved. Among them, 140 had Dynamic CTA with at least two phases of acceptable resolution. From those, 68 patients had both arterial and venous phase images, each including both the original post-contrast CT image as well as the bone and soft tissue subtracted image. 11 patients were randomly selected to evaluate the accuracy of the template. The retrospective study was approved by the IRB with a waiver of informed consent form.

\subsection{Preprocessing}

\begin{figure}
   \begin{center}
   \includegraphics[clip, trim=0in 1.2in 0in 1.2in,width=\textwidth]{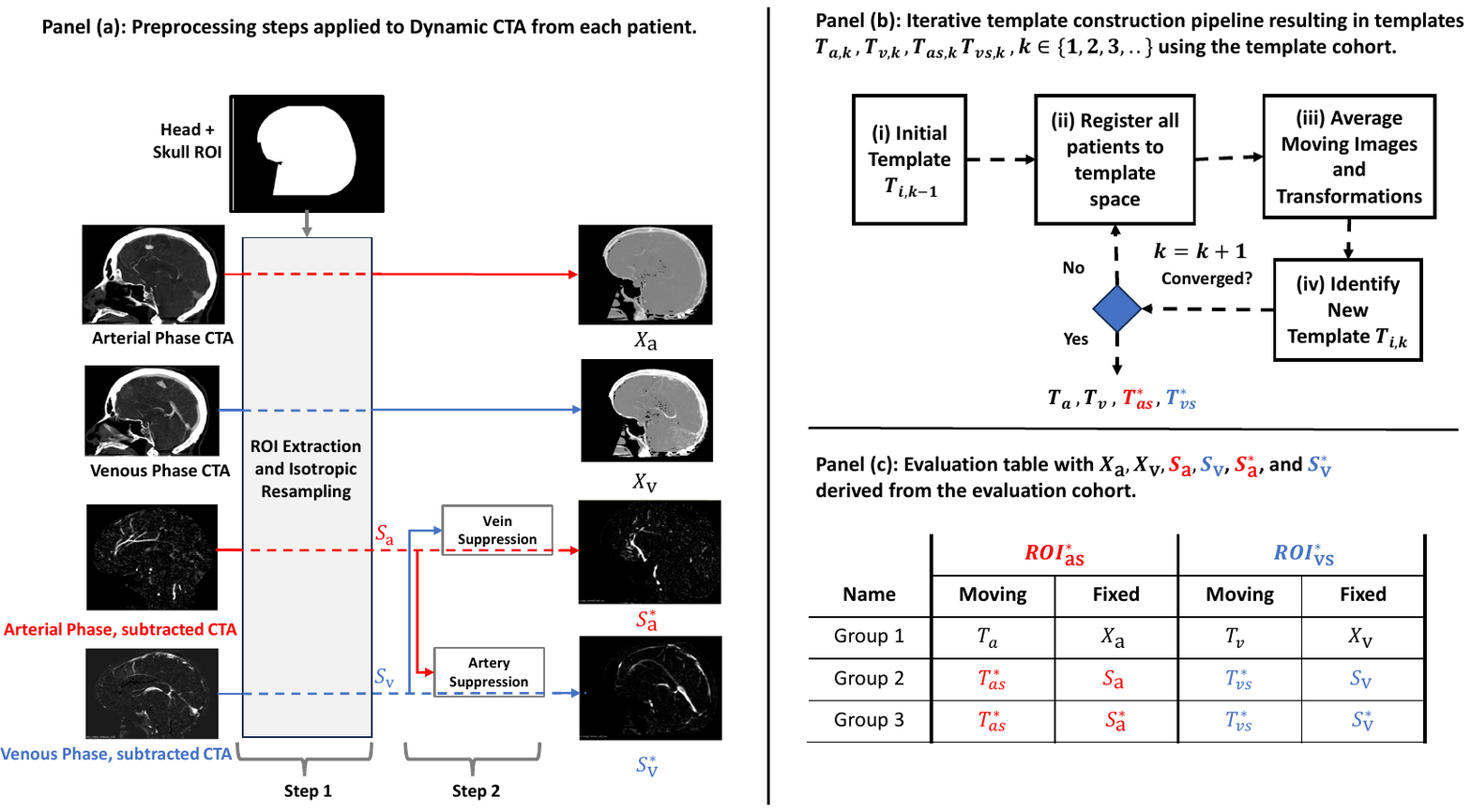}
   \caption
   {(a) Preprocessing steps for a dynamic CTA from a single patient. For both phases, a head ROI is used on the CTA and the subtracted CTA to suppress signals from outside the skull.  Next, the subtracted images are processed using Algorithm 1 to create the vessel-separated images. (b) Shows the workflow used to identify the optimal template. From all patients from the template cohort and for each group of images $X_a,X_v,S_a^*,$ and $S_v^*$, the pipeline creates intermediate templates $T_{a,k},T_{v,k},T_{as,k},$ and $T_{vs,k}$ respectively, at iteration $k$. After convergence, the identified optimal template is labelled as $T_{a},T_{v},T_{as}^*,$ and $T_{vs}^*$, respectively. (c) Shows how the evaluation of the registration-based segmentation is conducted with the evaluation cohort. For both arterial and venous ROIs, we test three groups of registrations with different combinations of moving images and fixed images. The identified non-linear transformation is then applied to the corresponding ROI.
   \label{fig:preprocessing} 
    }  
    \end{center}
\end{figure}

CT data were converted to NifTi format using the dcm2niix package\cite{li_first_2016}. Figure~\ref{fig:preprocessing} (a) illustrates the processing steps to create images ready for analysis. To suppress extracranial soft tissues, particularly of the neck, which are sometimes incompletely subtracted because of motion from breathing and atrial pulsation, we outlined a binary ROI of the skull and neck on a publicly available CT template\cite{muschelli_publicly_2020}. This CT template was affine registered to the CTA image, and the resulting transformation matrix was applied to the predefined ROI to create a binary mask of the skull and neck in the patient space. Any voxels outside of this ROI were excluded from further analysis. Lastly, all images were resampled to achieve an isotropic resolution of 0.468mm. 

Arterial and venous structures were separated into distinct volumes for each patient by applying Algorithm 1 to the subtracted CTA images. This algorithm was specifically designed  to selectively suppress venous structures in the arterial phase and arterial structures in the venous phase images. The inputs to the algorithm were the subtracted arterial ($S_a$) and venous ($S_v$) phase images, along with their corresponding CTA images ($X_a$ and $X_v$). Note that $S_a$ and $S_v$ are generated by the scanner (Figure~\ref{fig:preprocessing}(a)).  Rigid transformations, as described in Steps 1 and 2 of Algorithm 1, were implemented to align the images and minimize suppression artifacts caused by patient motion between acquisitions. Steps 4 and 5 of the algorithm subsequently enhanced the contrast-to-background ratio for arteries and veins, respectively. The outcome of applying Algorithm 1 for each patient are two distinct volumes, $S_a^*$ and $S_v^*$, representing bone and vein-suppressed arterial phase, and bone and artery-suppressed venous phase images, respectively.


\begin{algorithm}
\caption{Artery and Vein Suppression. \\ 
\textbf{Input:} $S_a$, $S_v$, $X_a$, $X_v$ \\ 
\textbf{Output:} $S^*_a$, $S^*_v$}
\begin{small} 
\footnotesize
    \begin{algorithmic}[1]
    \State Compute rigid transformation $G_{rv}$ such that $X_a$ is moving and $X_v$ is fixed.
    \State Compute rigid transformation $G_{ra}$ such that $X_v$ is moving and $X_a$ is fixed.
    \State $S_{v \rightarrow a} \gets G_{ra}(X_v)$, \text{Transform $X_v$ so that it is in the same space as $X_a$.} 
    \State $S_{a \rightarrow v} \gets G_{rv}(X_a)$, \text{Transform $X_a$ so that it is in the same space as $X_v$.}
    \State $s^*_a \gets \begin{cases} 
    0 & \text{if } s_a < s_{v \rightarrow a} \\
    s_a & \text{if } s_a > s_{v \rightarrow a} 
    \end{cases}$, \text{where $s^*_a$, $s_{v \rightarrow a}$, and $s_a$ are elements of the matrix $S^*_a$, $S_{v \rightarrow a}$, and $S_a$ respectively.}
    
    \State $s^*_v \gets \begin{cases} 
    0 & \text{if } s_v < s_{a \rightarrow v} \\
    s_v & \text{if } s_v > s_{a \rightarrow v} 
    \end{cases}$, \text{where $s^*_v$, $s_{a \rightarrow v}$, and $s_v$ are elements of the matrix $S^*_v$, $S_{a \rightarrow v}$, and $S_v$ respectively.}
    
    \end{algorithmic}
\end{small} 

\end{algorithm}

\subsection{Multivariate Template Creation}

Since the first published pipeline to create an imaging template\cite{avants_optimal_2010-1}, multiple studies\cite{ryan_template_2020,muschelli_publicly_2020} have utilized the iterative procedure. Figure~\ref{fig:preprocessing}  (b) provides an overview of the pipeline: 
In stage (i), the initial unbiased templates for the arterial phase CTA ($T_{a,0}$), venous phase CTA ($T_{v,0}$), subtracted arterial phase CTA ($T_{as,0}$), and subtracted venous phase CTA ($T_{vs,0}$) were identified use the same three steps: For each series, first, images are averaged across all patients. Second, the center of mass (CoM) of all patients in the series is aligned to this averaged image. Lastly, the average of these CoM-aligned images is computed. In stage (ii), for any iteration $k (k>0)$ and series $i$, all patients' images were registered to the existing template (denoted by $T_{i,k-1}$). Stage (iii) involves averaging both the transformations and the moved images. In the final Stage (iv), the new template $T_{i,k}$ is calculated and reused in Stage (ii).

Similar to the previously published head-CT template\cite{muschelli_publicly_2020}, convergence of the pipeline to the optimal template for each of the four series ($T_{a},T_{v},T_{as}^*,$ and $T_{vs}^*$) was established using the Dice Similarity Coefficient (DSC) to measure the overall change in the template's shape, and the root mean squared error (RMSE) to assess the change in voxel intensity between consecutive iterations ($k$ and $k-1$). Both metrics are calculated over a predefined mask rather than the entire volume. The mask used for calculating the DSC includes all voxels with a positive non-zero value ($T_{i,k}>0$) at iteration $k$. The mask for RMSE, at iteration $k$ is defined as the union of positive non-zero voxels from iterations $k$ and $k-1$ (i.e., $T_{i,k}>0 | T_{i,k-1}>0$, where $|$ represents the logical OR operator). 

As the desired template would require a high DSC and low RMSE, we set a DSC cutoff of 98\% and chose the iteration with the lowest RMSE. As described in Figure~\ref{fig:preprocessing}  (b), following convergence, the arterial phase CTA template ($T_{a}$), the venous phase CTA template ($T_{v}$), the vessel-separated artery template ($T_{as}^*$), and the vessel-separated vein template ($T_{vs}^*$) are created. 

The range of attenuation values in the angiographic templates were 0-375HU and 0-152HU for $T_{as}^*$ and $T_{vs}^*$. To extract the binary arterial and venous ROIs from templates, we used 3D-Slicer\cite{kikinis_3d_2014}. First, a preliminary segmentation was performed using a threshold of 30HU. Next, preserving the largest connected island eliminated noisy disconnected segments surrounding the arteries and veins. A radiologist then reviewed and corrected the final ROI. The Supplementary Figure~\ref{fig:reg_seg} shows the surface rendering of the identified arterial ($\text{ROI}_a^*$) and venous ROIs ($\text{ROI}_v^*$). The distal intracranial arteries and veins are not captured in the templates (and the ROI), likely due to their high spatial variability and  small sizes.

\subsection{Deep learning model training}

The CTA data was first split into validation and training sets. As the training set consisted of multiple CTA of the same patient, assignments to validation sets were performed on the patient level instead of the CTA level. Specifically, patient-level sampling was performed to prevent a single patient from appearing in validation and training sets. Following the guidelines of the training framework\cite{isensee_nnu-net_2021}, the default settings for both fixed and rule-based parameters were used. Each model fold was trained 1000 epochs with a linearly decaying learning rate starting with 0.01. The model parameters were optimized using stochastic gradient descent. The loss function was the average of dice loss and cross-entropy loss. For the rule-based parameters, the patch and batch sizes were 90x160x160 and two, respectively. Given the nature of the data, the CT normalization scheme was used. The models were trained on an NVIDIA GeForce RTX 4090 GPU. Details of the architecture can be found in the original manuscripts. In addition, we also train the NexToU\cite{shi_nextou_2023} model given the topological nature of the segmentation task and its strong performance of the TopCoW\cite{yang_topcow_2024} segmentation challenge.  

\begin{figure}
    \centering
    \includegraphics[clip, trim=0in 1.5in 0in 1.5in, width = \textwidth]{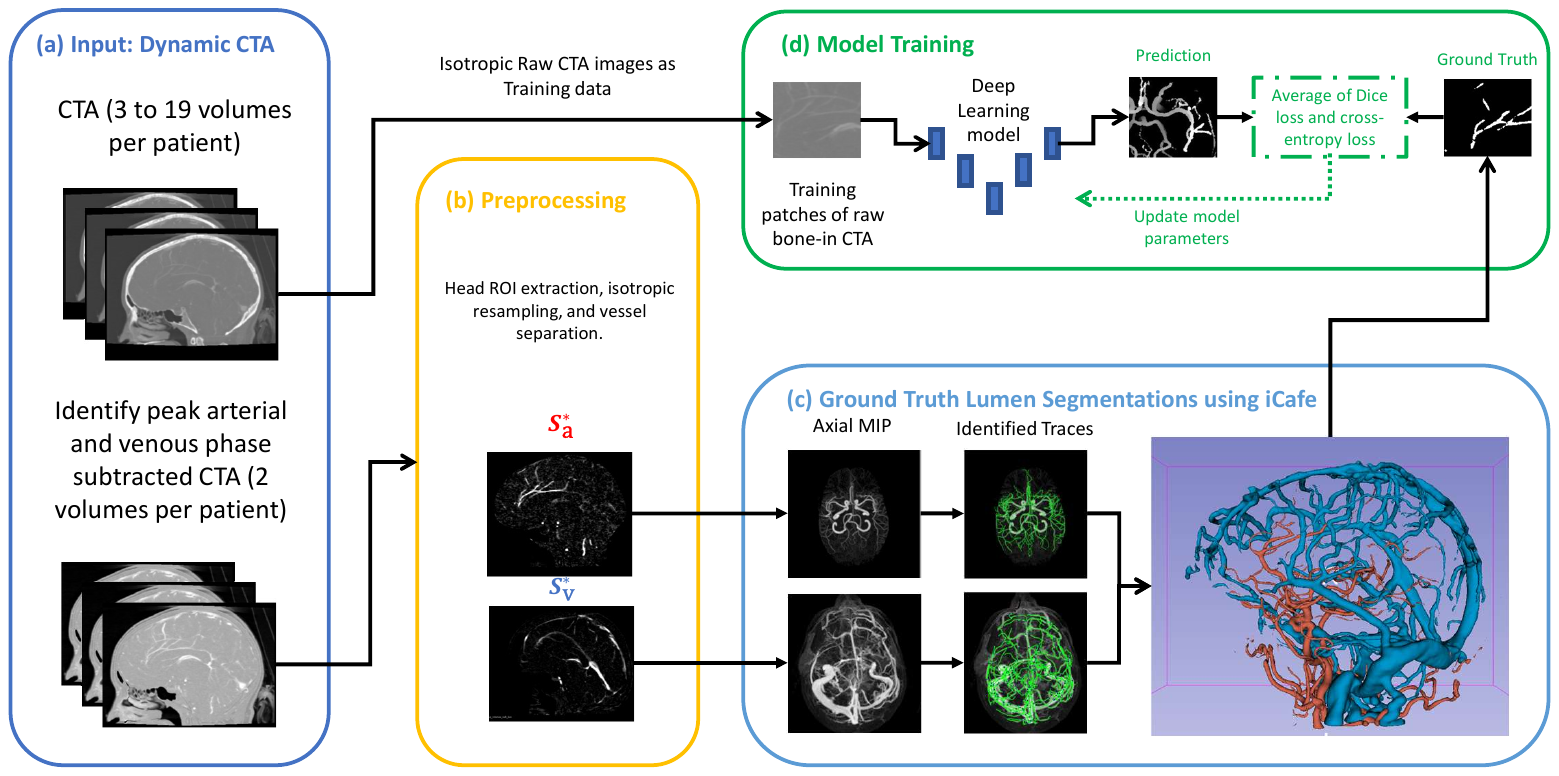}
    \caption{Dynamic CTA for all patients are collected (Step (a)). The subtracted images are processed to suppress voxels outside the head and resample to 0.468mm$^3$ isotropic resolution. Finally, the vessel-separated, artery-only ($S_a^*$) and vein-only ($S_v^*$) are computed (Step (b)). The vessel-separated images are then processed individually using iCafe's vessel tracing algorithm. These traces are then validated and exported for model training (Step (c)). In Step (d), for the model training, all available CTA data (i.e., 3-19 volumes per patient) is paired with the corresponding ground truth segmentation, and a deep learning model is trained.}
    \label{fig:model_training}
\end{figure}

\subsection{Ground Truth Segmentations}

\begin{figure} 
   \begin{center}
   \includegraphics[clip, trim=0in 3in 0in 3in,width=\textwidth]{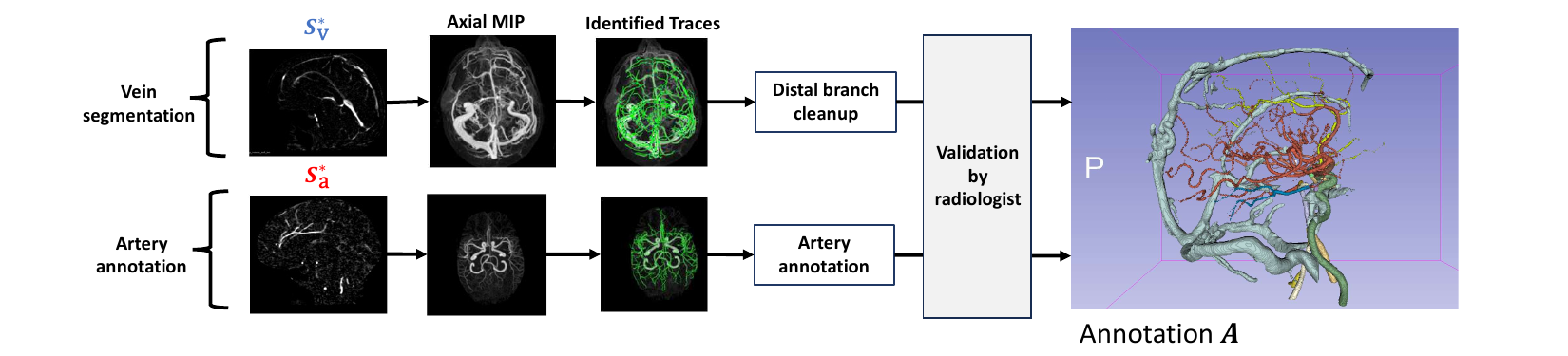}
   \caption{An overview of extracting the ground truth segmentations of veins and arteries using iCafe. Traces are first identified for both vessels. For the artery-only volume, the traces are edited for accuracy and exported into segmentations. For the veins, the vessel traces are directly exported into segmentations. Distal branches from the venous segmentation were removed using 3D Slicer. \label{fig:gt_annotation_workflow} 
    }  
    \end{center}
\end{figure}

To create ground truth ROIs on the evaluation dataset described in Figure~\ref{fig:gt_annotation_workflow} Panel (c) was used. The preprocessing steps in \ref{fig:preprocessing} (a) were used to create the vessel-separated volumes $S_a^*$ and $S_v^*$ from the evaluation cohort. To segment the arteries and veins from $S_a^*$ and $S_v^*$, we used iCafe's feature extraction pipeline. This application uses intensity normalization followed by rudimentary segmentation using Phansalkar threshold\cite{phansalkar_adaptive_2011} (used for vessel tracing) and the Renyi entropy threshold\cite{kapur_new_1985} (used for visualization). To identify vessel-like structures, the tool uses a vessel tracing algorithm based on a modified open-curve active contour model\cite{wang_broadly_2011}. Spurious vessel traces outside the skull and in the soft tissue of the head are deleted manually.

\subsubsection{Artery Ground Truth Segmentation} 

After creating the vessel traces, the arteries were labelled using iCafe with two modifications of the recommended approach for MRA vessel annotation\cite{chen_development_2018}: (1) The ground truth segmentations included the proximal and first-order arteries. The second and third-order arteries are labeled jointly as ``M2+'', ``P2+'', and ``A2+'' for the MCAs, PCAs, and ACAs. They were grouped for two reasons: (a) The complexity of identifying the M2 and M3 junctions, and (b) the third-order arteries of ACA and the PCA did not have a dedicated artery label. (2) Because iCafe did not use an artery label for the early cortical branches of MCAs, they were labeled as ``M2+''. Therefore, the final segmentations dataset contained the following 20 artery annotations: ``ICA (Left)'', ``ICA (Right)'', ``VA (Left)'', ``VA (Right)'', ``BA'', ``PComm (Left)'', ``Pcomm (Right)'', ``P1 (Left)'', ``P1 (Right)'', ``P2+ (Left)'', ``P2+ (Right)'', ``M1 (Left)'', ``M1 (Right)'', ``M2+ (Left)'', ``M2+ (Right)'', ``A1 (Left)'', ``A1 (Right)'', ``A2+ (Left)'', ``A2+ (Right)'', and ``Acomm''. All annotations for the evaluation dataset were exported into 3D volumes segmentations of the vascular structures. These annotations were validated by a subspecialty-trained neuroradiologist by overlaying the segmentations with the unprocessed subtracted images $S_a$ in 3D Slicer (https://www.slicer.org). Inaccuracies in annotation due to anatomical variability in the arteries were corrected.

\subsubsection{Venous Ground Truth Segmentation}

Since iCafe lacks a dedicated venous annotation pipeline, the ground truth vessel traces of the veins for the evaluation dataset were exported. Then, 3D Slicer was used by the neuroradiologist to remove the distal branches. In the final venous segmentation, superior sagittal sinus, straight sinus, transverse sinus, sigmoid sinus, and cavernous sinus were incorporated into a single venous label volume. Given the clinical importance of the cavernous sinuses (CVS), these structures were specifically segmented into a distinct volume within the venous segmentations. This additional step ensures accurate delineation and introduces a new second venous label for the CVS.

The Supplementary Figure~\ref{fig:gt} shows the surface rendering of the arterial and venous annotations, with each arterial branch represented in a different color. To account for patient motion between the acquisition of $X_a$, $X_v$, $S_a$, and $S_v$, all series were rigid registered to the arterial phase CTA ($X_a$). This rigid transformation was also applied to the annotated arterial and venous annotations initially identified in the same space as $S_a$ and $S_v$. These manually segmented, $X_a$-aligned labels (20 arterial and 2 venous) were used as ground truth (GT)  for quantitative evaluation of our templates. 

\subsection{Evaluation}

\begin{figure} 
   \begin{center}
   \includegraphics[clip, trim=0in 1in 0in 4.5in,width=\textwidth]{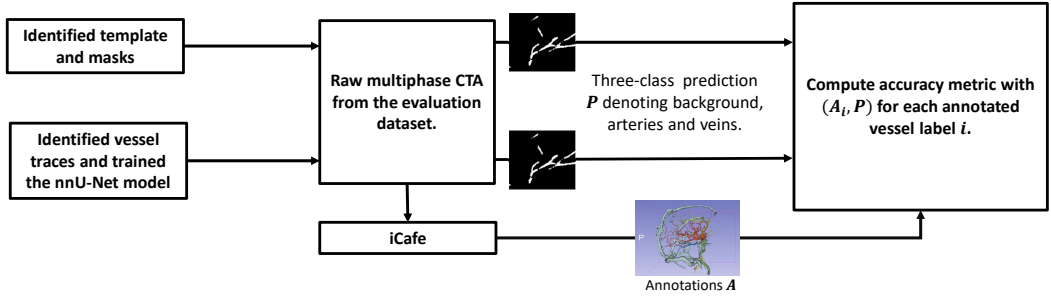}
   \caption{Overview of the evaluation procedure. We register angiographic CT template to the multiphase CTA data from the evaluation cohort. This gives us a three-class segmentation (background, arteries, veins). In parallel, vessel traces annotated via iCafe and trained on nnU-Net provide an additional segmentation result. Both methods are assessed against ground-truth labels to compute accuracy metrics for each annotated vessel label. \label{fig:eval_criteria} 
    }  
    \end{center}
\end{figure}

Using the ground truth segmentations identified in the previous subsection, we describe the our approach to validating the registration- and learning- based segmentation from the template and the deep learning models respectively. Figure~\ref{fig:eval_criteria} shows an overview of the process. In general, the registration- and learning-based segmentation result in a volume with three classes. The ground truth segmentations derived from iCafe is then used to compute the accuracy metrics for each of the predictions $\textbf{P}$ from both approaches. We now describe the approach for evaluation in both methods in detail.

\subsubsection{Template ROI Validation}

To evaluate the accuracy of the ROIs ($\text{ROI}_a^*$ and $\text{ROI}_v^*$) from our templates and to quantify the alignment accuracy of the templates, we registered the templates to individual patient images in the evaluation cohort. As shown in Figure~\ref{fig:preprocessing} Panel (c), three different groups of registrations were performed: ``Group 1'': Registration of CT templates ($T_a$ and $T_v$) to the patient's corresponding arterial and venous phase unsubtracted CTA ($X_{a}$ and $X_{v}$). ``Group 2'': Register the angiographic templates ($T_{as}^*$ and $T_{vs}^*$) to the patient's corresponding subtracted images ($S_{a}$ and $S_{v}$). And ``Group 3'': Registration of the angiographic templates ($T_{as}^*$ and $T_{vs}^*$) with the patient's corresponding subtracted and vessel-separated volumes ($S_a^*$ and $S_v^*$).

These three resulting transformation matrices were applied to  $\text{ROI}_a^*$ and $\text{ROI}_v^*$, transforming the ROIs from our templates to each patient's anatomic space. The vessels corresponding to the transformed $\text{ROI}_a^*$ and $\text{ROI}_v^*$ were labeled `artery' and `vein' respectively. 

We evaluated the vessel segmentation accuracy using a modified Dice coefficient (mDC) to emphasize sensitivity. Our modified coefficient (unlike the traditional Dice similarity coefficient) uses only the GT segmentation in the denominator. Specifically, let volume $A_j$ represent the vessel mask of interest ($j=1, \ldots, 22$ for 20 arterial and 2 venous labels), and volume $P$ be the predicted binary label for the arterial or venous ROI. The modified Dice coefficient is defined as: $\text{mDC} (A_j, P) = \frac{|A_j \cap P|}{|A_j|}$.

\subsubsection{Deep learning model Validation}

Choosing the correct evaluation metric is critical for vessel segmentation\cite{yang_topcow_2024}. Consequently, following established guidelines from the segmentation challenge, we evaluate the model with counting, distance, and centerline-based metrics. As the predictions for both the nnU-net and NexToU models predict three classes, all three metrics were modified to focus on the sensitivity of the vessel segmentation. Figure 3, Panel (b) provides an overview of the inference workflow.  Let volume $A_i$ be the vessel label with ``1'' indicating a vessel branch $i$ of interest, and ``0'' indicating background, and volume $P$ the predicted label with ``1'' indicating either arteries or veins and ``0'' denoting background. The segmentations were evaluated using three metrics:

\begin{enumerate}
    \item \textbf{Modified dice coefficient}: The dice coefficient is a commonly used metric for evaluation segmentations. The modified Dice coefficient focusing on sensitivity was defined as:
    \[
    mDC(A_i,P) = \frac{|A_i \cap P|}{|A_i|}
    \]
    
    \item \textbf{Average directed Hausdorff surface Distance}: The directed variation of the Hausdorff distance was chosen based on a ranked comparison of multiple distance metrics in the context of vessel segmentation\cite{aydin_evaluation_2021}. While a subsequent effort identified a better directed metric Balanced Housdorff distance\cite{aydin_usage_2021}, we limit the metric calculation to the first term in the metric equation. This is because, in our application, the boundary of the predicted volume is significantly larger than the ground truth. Further, to reduce the computational load, the distances were calculated on the segmentation surface rather than the whole volume. Consequently, the metric $adHD_i$ was defined as:
    \[
    adHD(A_{s,i}, P_s) = \frac{1}{|A_{s,i}|} \sum_{p \in A_{s,i}} d(p, P_s)
    \]
    where $d(p, P_s) = \min_{p' \in P_s} |p - p'|$, and $A_{s,i}$ and $P_s$ are the surfaces of the structures in volumes $A_i$ and $P$ respectively.
    
    \item \textbf{Centerline metric}: Finally, we compute topology sensitivity ($tSens_i$) to quantify the extent of centerlines that are captured in the vessel segmentation model. For completeness, this metric is defined based on the \textit{clDice}\cite{shit_cldice_2021} metric as:
    \[
    tSens(A_{c,i}, P) = \frac{|A_{c,i} \cap P|}{|A_{c,i}|}
    \]
    where $A_{c,i}$ is the centerline volume of $A_i$.
\end{enumerate}

Finally, as both models were trained on all available contrasts of the Dynamic CTA cohort, the models were also evaluated on three distinct phase images: 
\begin{enumerate}
    \item Arterial Phase: This was used to assess the model performance when the arteries are the brightest,
    \item Venous Phase: This was used to assess the model performance when the veins are brighter than the arteries and
    \item No-Contrast Phase: Dynamic CTA exams in the evaluation dataset with the first volume without visible contrast were flagged as the no-contrast phase. This was used to assess model performance where no contrast information was available.
\end{enumerate}




\section{Results}

\label{sec:results}

\begin{table}
\begin{center}
\caption{Demographic characteristics of both the training and and evaluation datasets. 
\label{table1}
\vspace*{2ex}
}
\begin{tabular} {|c|c|c|}
\hline
Cohort (Size) & Age in Years (25\%, Mean, 75\%, Range) & Gender (F, M) \\ 
\hline
Training (27) & 40, 52, 63, 19-75 & 17, 10 \\ 
\hline
Evaluation (11) & 58, 61, 68, 42-74 & 11, 0 \\ 
\hline
\end{tabular}
\end{center}
\end{table}

\begin{figure} 
   \begin{center}
   \includegraphics[width=8cm]{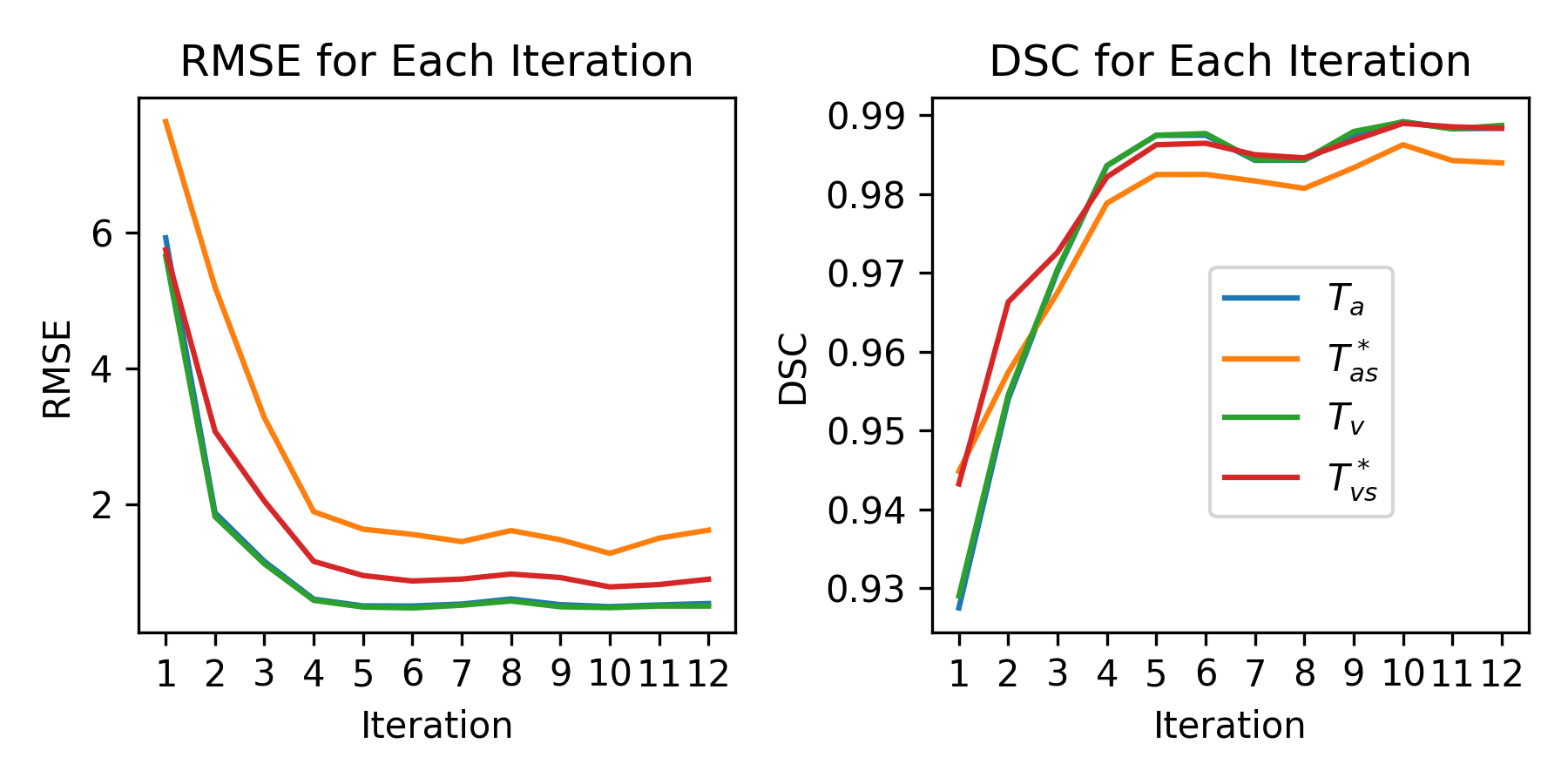}
   \caption{RMSE and DSC for each iteration in the template construction pipeline for all images $T_{a} (\text{blue}), T_{v}(\text{green}), T_{as}^*(\text{orange})$, and $T_{vs}^*(\text{red})$.
   \label{fig:convergence} 
    }  
    \end{center}
\end{figure}

\begin{figure} 
   \begin{center}
   \includegraphics[width=\textwidth]{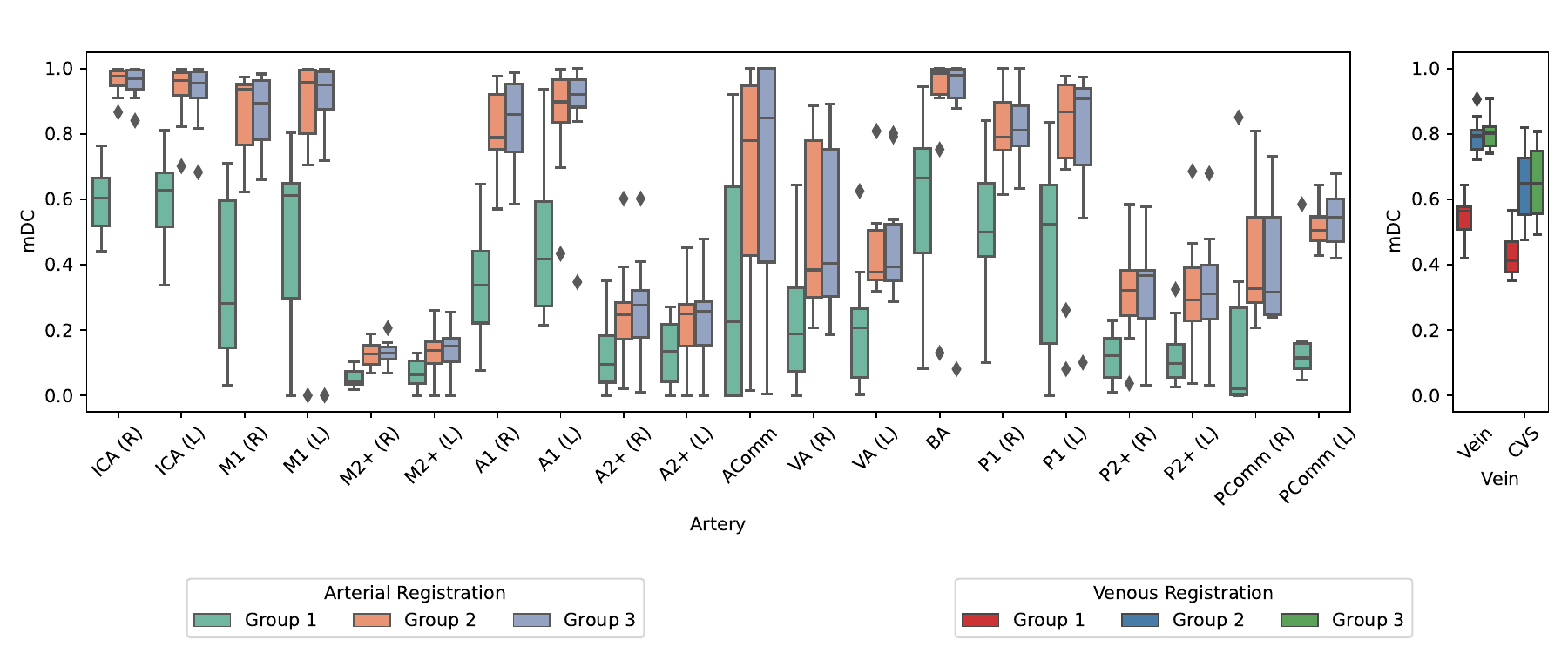}
   \caption{Modified dice coefficient (mDC) for each arterial and venous label. Three registrations are tested for each artery and vein label (on the x-axis).  
   \label{fig:segmetnation_accuracy} 
    }  
    \end{center}
\end{figure}

\begin{table}
\begin{center}
\scriptsize
\caption{Comparing different phases for nnU-Net and NexToU}
\label{tab:comparison}
\vspace*{2ex}
\begin{tabular} {|c|c|c|c|c|c|c|c|c|}
\hline
\textbf{Artery Order} & \multicolumn{2}{c|}{\textbf{No Contrast}} & \multicolumn{2}{c|}{\textbf{Arterial Phase}} & \multicolumn{2}{c|}{\textbf{Venous Phase}} & \textbf{nnU-Net Avg.} & \textbf{NexToU Avg.} \\ 
\cline{2-7}
& \textbf{nnU-Net} & \textbf{NexToU} & \textbf{nnU-Net} & \textbf{NexToU} & \textbf{nnU-Net} & \textbf{NexToU} & & \\ 
\hline
\multicolumn{9}{|c|}{\textbf{mDC}} \\ 
\hline
Proximal & 0.53 & 0.47 & 0.96 & 0.97 & 0.89 & 0.90 & 0.79 & 0.78 \\ 
\hline
First order & 0.56 & 0.54 & 0.97 & 0.97 & 0.94 & 0.94 & 0.82 & 0.82 \\ 
\hline
Second order & 0.12 & 0.09 & 0.76 & 0.77 & 0.59 & 0.59 & 0.49 & 0.48 \\ 
\hline
Vein & 0.68 & 0.65 & 0.76 & 0.74 & 0.86 & 0.85 & 0.77 & 0.75 \\ 
\hline
Average & 0.47 & 0.44 & 0.86 & 0.86 & 0.82 & 0.82 & 0.72 & 0.71 \\ 
\hline
\multicolumn{9}{|c|}{\textbf{tSens}} \\ 
\hline
Proximal & 0.62 & 0.57 & 0.99 & 0.99 & 0.97 & 0.97 & 0.86 & 0.84 \\ 
\hline
First order & 0.64 & 0.64 & 0.99 & 0.99 & 0.98 & 0.98 & 0.87 & 0.87 \\ 
\hline
Second order & 0.06 & 0.04 & 0.60 & 0.61 & 0.40 & 0.41 & 0.35 & 0.35 \\ 
\hline
Vein & 0.53 & 0.50 & 0.63 & 0.62 & 0.76 & 0.75 & 0.64 & 0.62 \\ 
\hline
Average & 0.46 & 0.44 & 0.80 & 0.80 & 0.78 & 0.78 & 0.68 & 0.67 \\ 
\hline
\multicolumn{9}{|c|}{\textbf{adHD (mm)}} \\ 
\hline
Proximal & 1.90 & 3.03 & 0.05 & 0.04 & 0.17 & 0.21 & 0.71 & 1.09 \\ 
\hline
First order & 1.00 & 1.22 & 0.03 & 0.02 & 0.06 & 0.05 & 0.36 & 0.43 \\ 
\hline
Second order & 11.67 & 16.99 & 1.57 & 1.61 & 3.11 & 3.16 & 5.45 & 7.25 \\ 
\hline
Vein & 2.23 & 2.69 & 1.28 & 1.50 & 0.53 & 0.59 & 1.35 & 1.59 \\ 
\hline
Average & 4.20 & 5.98 & 0.73 & 0.80 & 0.97 & 1.00 & 1.97 & 2.59 \\ 
\hline
\end{tabular}
\end{center}
\end{table}

Table 1 presents the demographic characteristics of the two cohorts used for template creation and evaluation. The mean ages for the template construction and  evaluation datasets were 52 and 61, respectively. The age range for the construction cohort were 19-75, while  it was 42-74 for the evaluation cohort. Additionally, about two-thirds of the patients in the template construction cohort were female, whereas the evaluation cohort was entirely female. Of the 36 patients, 22 had meningiomas and 12 had glioblastomas, schwannomas, and fibrous tumors. In 19 of these patients,  cancer did not affect the cerebral vasculature. In 8 patients, cancer displaced the arteries without causing narrowing, in 9 patients, veins were either occluded or severely narrowed by tumors.

\subsection{Vessel template construction}

Figure~\ref{fig:convergence} illustrates the convergence of the metrics RMSE and DSC for the four templates. The convergence of CTA templates $T_a$ and $T_v$ represented by the blue and green lines,  are visually identical. Furthermore, the RMSE and DSC converge by iteration 5 and 10. As Constructing the template was computationally demanding, the pipeline was run for 12 iterations before identifying the final templates. The template at iteration 10 had the least RMSE and was chosen as the final image.

\subsection{Registration based segmentation performance}

Figure~\ref{fig:segmetnation_accuracy} shows the accuracy of registration-based segmentation for all labels. The three groups of registrations are shown in three box plots for arteries and veins.  The average mDC for Group 1 registrations in the proximal and first-order arteries (comprising of ICA, VA, BA, M1, A1, P1, and the communicating arteries PComm and AComm) was 0.39 mDC, and for the vein and CVS registration, 0.49 mDC. In Group 2, the average segmentation performance improved to 0.69 mDC for the proximal and first-order arteries and 0.72 mDC for the vein and CVS. In Group 3, the average segmentation performance was 0.72 mDC for the proximal and first-order arteries. The vein and the CVS average at 0.73 mDC. The distal vessels are not captured accurately, regardless of the registration group, as demonstrated by an average of 0.12 mDC in M2+, A2+, and P2+.

Supplementary Figure~\ref{fig:good_bad_ex} shows one instance where the ICA was accurately segmented and a second in which the ICA was poorly segmented. The ICA segmentation is the least accurate in the inferior sections near the skull base and the upper neck. Segmentation was worse for more tortuous vessels. Pair-wise statistical comparison of Groups 1, 2, and 3 confirmed that registering to the subtracted (Group 2) and vessel separated (Group 3) target images resulted in significantly better registration-based vessel segmentation compared to arterial and venous phase CTA images (Group 1). There was no significant difference between Group 2 (vessel-only template) and Group 3 (isolated artery and vein templates) for most vessels (Supplemental Table~\ref{tab:method1_method2_vs_method3}). 

\subsection{Deep learning model segmentation performance}

The three segmentation metrics (\textit{mDC}, \textit{adHD}, and \textit{tSens}) were assessed for different vessel groups: proximal, first-order, and second-order/distal veins. Generally, all metrics showed improvement from the no-contrast to the arterial phase and maintained high levels in the venous phase. For proximal vessels, the highest mDC was observed in the arterial phase for nnU-Net and NexToU, both at 0.96, while the lowest mDC was seen in the no-contrast phase for NexToU at 0.47. The tSens for proximal vessels was highest in the arterial phase, reaching 0.99 for both nnU-Net and NexToU, whereas the lowest tSens was in the no-contrast phase for NexToU at 0.57. The adHD for proximal vessels showed the best performance in the arterial phase with nnU-Net at 0.05 mm, and the worst in the no-contrast phase for NexToU at 3.03 mm.

First-order vessel segmentation metrics improvements. The highest mDC for first-order vessels was in the arterial phase for both nnU-Net and NexToU, both reaching 0.97. The lowest mDC was in the no-contrast phase for NexToU at 0.54. For tSens, the highest value was in the arterial phase at 0.99 for both nnU-Net and NexToU, while the lowest was in the no-contrast phase for NexToU at 0.64. The adHD was lowest in the arterial phase for nnU-Net at 0.03 mm, while the highest adHD was in the no-contrast phase for NexToU at 1.22 mm. Second-order and distal vessels, including veins, showed the most variability. The highest mDC was seen in the arterial phase for P2+ vessels with nnU-Net at 0.76, while the lowest was in the no-contrast phase for NexToU at 0.09. The tSens was highest for nnU-Net in the arterial phase at 0.60, and lowest for NexToU in the no-contrast phase at 0.04. The adHD was lowest in the arterial phase for nnU-Net at 1.57 mm, and highest in the no-contrast phase for NexToU at 16.99 mm. For veins, the adHD decreased significantly from 2.23 mm in the no-contrast phase to 0.53 mm in the venous phase with nnU-Net, while mDC and tSens improved from 0.68 to 0.86 and 0.53 to 0.76, respectively.

\section{Discussion}
\label{sec:discussion}
\subsection{Vessel templates construction}
We report the development of an atlas consisting of arterial and venous anatomical templates representing intracranial CTA vascular anatomy and the use of these templates for automated labeling of proximal and first-order arteries and major venous structures on intracranial CTA. These methods can contribute significantly to the development of vascular image analysis methods for assisting expert interpretation of intracranial CTA in patients with acute stroke, intracranial aneurysm, and other neurovascular disorders. Our pipeline exploits the high temporal resolution of dynamic 4D-CTA to overcome two stubborn challenges that have impeded the creation of angiographic atlases in CTA: bone subtraction and artery-vein separation. The high temporal resolution of the 4D-CTA images made bone and soft-tissue subtraction possible with minimized motion-related artifacts. We further exploited the time-resolved 4D-CTA images to suppress phase-dependent non-vascular structures and separate arteries and veins into two distinct volumes. Arterial and venous anatomic ROIs were derived from the combined angiographic artery and vein templates using a simple threshold. This represented large arteries (ICA, BA, M1, P1, and A1). Smaller distal vessels appeared to be diffused. 

Evaluating the accuracy of registration-based large vessel segmentation, required producing ground truth segmentation for the test set. The near-perfect background suppression and vessel separation allowed using software developed for MRA processing (iCafe).  Automatic vessel labeling allowed us to analyze the segmentation quality for individual arteries: left and right ICA, A1, M1, P1, posterior communicating artery, vertebral artery, basilar artery, and distal arteries (denoted as A2+, M2+, P2+). The lower mDC scores for distal and posterior communicating artery segmentations may be related to the greater variability of these structures and/or their small calibers. Visual inspection of registration to the unsubtracted CTA source images (Group 1) suggested that the registrations were driven mainly by the alignment of bones. This was in contrast to registering the template to the bone and soft tissue subtracted volumes (Group 2) or vessel-separated volumes (Group 3), which demonstrated a high mDC of 0.85 for proximal vessels, including the ICA, A1, M1, P1, and the BA, and 0.81 for veins.  

\subsection{Deep learning-based segmentation}

A deep learning model is developed for automatic CTA vessel segmentation that effectively distinguishes between arteries and veins. By leveraging Dynamic CTA to (1) subtract bones and soft tissues and (2) separate arterial and venous structures into different volumes, we addressed several critical challenges in vessel segmentation, including anatomical complexity, arterial-venous separation, and limited training data. iCafe, initially developed for MRA, significantly reduced the annotation burden. Next, Dynamic CTA allows us to incorporate CTA images reconstructed at different phases during training, thus exposing the model to a wider range of contrasts without creating new training examples. These factors allow for accurate segmentation with two deep learning models based on the nnU-Net framework and a dataset with 102 training images from 27 patients.

The segmentation quality was thoroughly assessed across three phases (No Contrast, Arterial Phase, and Venous Phase) using metrics that focused on sensitively. These were modified versions of the Dice coefficient (mDC), true sensitivity (tSens), and average directed Hausdorff Distance (adHD). Proximal vessels, including the ICA, A1, M1, and vertebrobasilar arteries, exhibited high segmentation quality with the highest mDC observed in the arterial phase for both nnU-Net and NexToU (0.96), while the no-contrast phase recorded the lowest mDC. The arterial phase also achieved the highest tSens values (0.99 for both models), contrasting with the lowest tSens in the no-contrast phase for NexToU (0.57). The best adHD was observed in the arterial phase with nnU-Net (0.05 mm), while the worst was in the no-contrast phase for NexToU (3.03 mm). First-order vessel segmentation mirrored these trends, with the highest mDC in the arterial phase (0.97 for both methods) and the lowest in the no-contrast phase for NexToU (0.54). Similarly, the arterial phase showed the best tSens (0.99 for both models), and the no-contrast phase had the lowest for NexToU (0.64). The adHD was lowest in the arterial phase for nnU-Net (0.03 mm) and highest in the no-contrast phase for NexToU (1.22 mm).

Second-order and distal vessels, including veins, displayed greater variability. The highest mDC for these vessels was observed in the arterial phase with nnU-Net (0.76), and the lowest in the no-contrast phase for NexToU (0.09). The highest tSens was in the arterial phase for nnU-Net (0.60), while the lowest was in the no-contrast phase for NexToU (0.04). The adHD was lowest in the arterial phase for nnU-Net (1.57 mm) and highest in the no-contrast phase for NexToU (16.99 mm). Vein segmentation showed significant improvements from the no-contrast to the venous phase: adHD decreased from 2.23 mm to 0.53 mm with nnU-Net, mDC increased from 0.68 to 0.86, and tSens improved from 0.53 to 0.76. These findings underscore the efficacy of dynamic CTA and our segmentation model in delivering accurate and robust vessel annotations across varying contrast phases.
The implications of these findings are substantial. The improved segmentation quality, especially in the arterial phase, highlights the potential of dynamic CTA in enhancing the accuracy of vascular imaging in conventional CTA. This ultimately can help reducet he cognitive burden on radiologists interpreting vascular abnormalities. Furthermore, the robust performance of the model across different contrast phases demonstrates its potential application in clinical scenarios and as prior information for deep learning models that focus on the vascular structures. 

\subsection{Limitations}

While the near-perfect background subtraction and artery-vein separation offered by dynamic 4D-CTA were critical to this work, they also impose significant restrictions. Specifically, the small size of our dynamic 4D-CTA dataset (25 patients for template construction, 11 for evaluation) stemmed from the limited availability of scans featuring optimal arterial and venous phase images. As a result, cases with poor timing or low-quality subtraction were excluded, thereby reducing the diversity of vascular anatomies.

\subsubsection{Limitations with this anatomical template} As the registration-based segmentation nessisitated the need for subtracted images available from dynamic CTA  data, this constrains the template’s applicability in settings with less advanced CT scanners. This arterial and venous template may not capture the breadth of anatomical variability, particularly regarding distal vessels in the MCA, PCA, and ACA territories, as well as the vertebral arteries. The confined sample size and the need for peak arterial and venous phases restricted the atlas to mainly proximal vessels, with the cavernous sinuses (CVS) requiring manual validation due to their atypical morphology. Future expansions of the 4D-CTA dataset, including more anatomically diverse cohorts, are planned to enhance coverage of distal vessels and to incorporate automated recognition of challenging venous structures such as the CVS.

\subsubsection{Limitations with vessel annotations}
Despite strong segmentation performance for proximal arteries and major veins, our learning-based approach still faces hurdles. First, the venous ground truth masks required partial manual editing in 3D Slicer due to the proximity of the cavernous sinuses to bony structures and the ICAs. Second, the iCafe annotation tool used for vessel tracing was originally designed for MRA and could benefit from specialized CT-focused enhancements that handle background noise and fine distal vessels more effectively. Third, although we demonstrated promising results on high-resolution images, the robustness of these deep learning models remains untested on lower-resolution data, which may be more common in urgent or resource-limited clinical environments. Finally, while enlarging the training dataset and improving semantic vessel labels are ongoing efforts, more extensive validations—potentially including inter-rater studies—are needed to solidify confidence in the model’s generalizability to a wider range of pathologies and scanner settings.

    

    

\section{Conclusion}
\label{sec:conclusion}

CTA images are a widely used imaging modality in the clinic. As it is used for diagnosis of emergent life-threatening conditions such as strokes and aneurysms, can greatly CT image analysis tools are critical for reducing the cognitive burden in radiologists interpreting these images. In this work, we leveraging Dynamic CTA for advancing conventional CTA. First, we construct dedicated arterial and venous CT atlases from dynamic 4D-CTA. We demonstrate, for the first time, that angiographic templates can reasonably isolate major vascular structures using non-linear registrations. Further, we also introduced an automatic vessel segmentation pipeline that addressed three challenges: anatomical complexity, arterial-venous separation, and limited training data. By leveraging bone-subtracted, vessel-separated images derived from dynamic 4D-CTA, our framework significantly reduces the annotation burden through the semi-automatic iCafe tool, originally designed for MRA. Moreover, our deep learning models achieved promising results on a modest dataset. This indicates that further expansion, such as including CT-specific adaptations of the annotation tool, a more diverse patient population, and additional validation of distal venous structures could expand the clinical impact of dynamic CTA by improving the clinical utility of conventional CTA. Overall, these findings underscore the potential of dynamic CTA and modern segmentation methods to advance intracranial vascular imaging, thereby enhancing clinical decision-making and patient care.

\bibliography{./references}      



\bibliographystyle{./medphy.bst}    



\end{document}


\title{Supporting Material}
\author{}

\maketitle

\clearpage

\begin{figure}[t]
    \centering
    \begin{minipage}[b]{0.19\textwidth}
        \subcaption{$X_{a}$}
        \includegraphics[clip, trim=3in 38in 3in 12.5in, width=\textwidth, height=0.9\textheight, keepaspectratio]{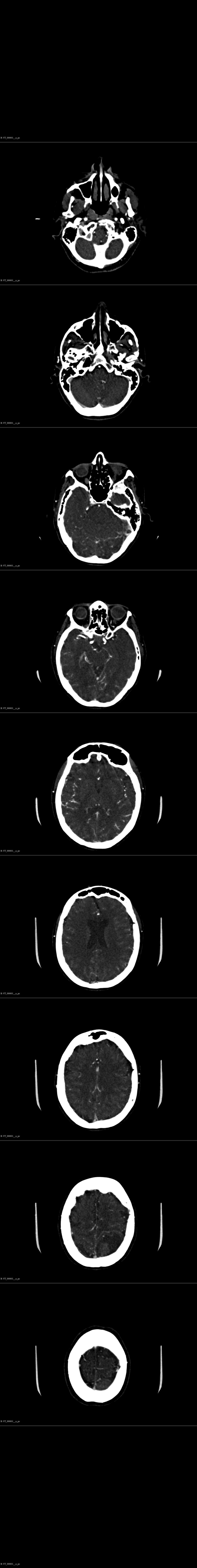}
    \end{minipage}
    \begin{minipage}[b]{0.19\textwidth}
        \subcaption{$S_{a}$}
        \includegraphics[clip, trim=3in 38in 3in 12.5in, width=\textwidth, height=0.9\textheight, keepaspectratio]{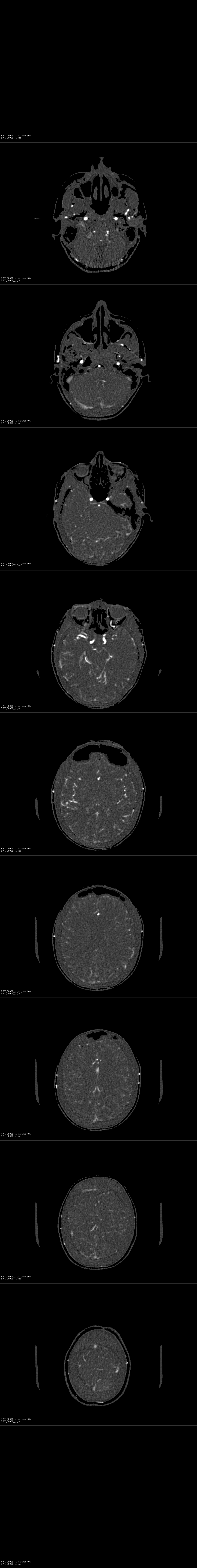}
    \end{minipage}
    \begin{minipage}[b]{0.19\textwidth}
        \subcaption{$S_{a}^*$}
        \includegraphics[clip, trim=3in 38in 3in 12.5in, width=\textwidth, height=0.9\textheight, keepaspectratio]{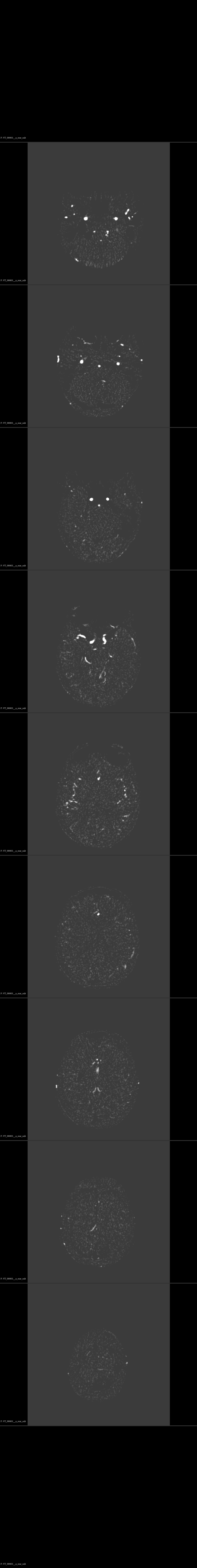}
    \end{minipage}
    \begin{minipage}[b]{0.19\textwidth}
        \subcaption{$S_{v}$}
        \includegraphics[clip, trim=3in 38in 3in 12.5in, width=\textwidth, height=0.9\textheight, keepaspectratio]{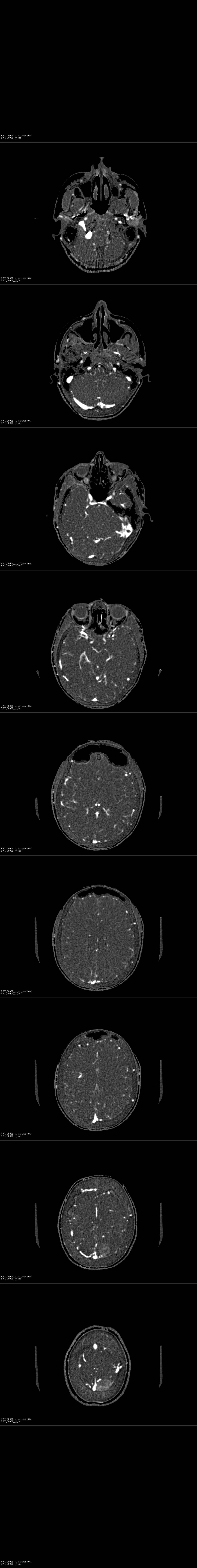}
    \end{minipage}
    \begin{minipage}[b]{0.19\textwidth}
        \subcaption{$S_{v}^*$}
        \includegraphics[clip, trim=3in 38in 3in 12.5in, width=\textwidth, height=0.9\textheight, keepaspectratio]{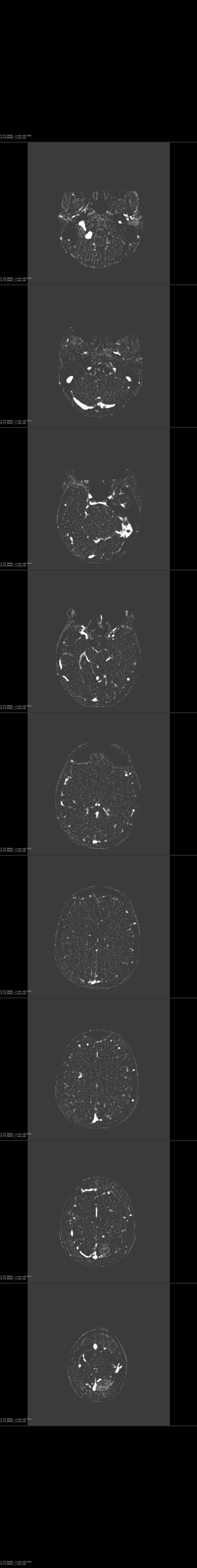}
    \end{minipage}
    \caption{Axial images of arterial phase CTA ($X_{a}$), bone subtracted arterial and venous phase ($S_{a}$ and $S_{v}$), and the vessel separated arterial and venous phase ($S_{a}^*$ and $S_{v}^*$) volumes in a sample patient.}
    \label{fig:aligned_strips}
\end{figure}
\FloatBarrier

\begin{table}[htbp]
\begin{center}
\captionv{10}{}{Demographic characteristics of both the template and the evaluation datasets. 
\label{tab_label}
\vspace*{2ex}
}
\begin{tabular} {|c|c|c|}
\hline
Cohort (Size) & Age in Years (25\%, Mean, 75\%, Range) & Gender (F, M) \\ 
\hline
Template (25) & 40, 52, 66, 19-75 & 16, 9 \\ 
\hline
Evaluation (11) & 58, 61, 68, 42-74 & 11, 0 \\ 
\hline
\end{tabular}
\end{center}
\end{table}

\begin{figure}[ht]
   \begin{center}
   \includegraphics[width=\textwidth]{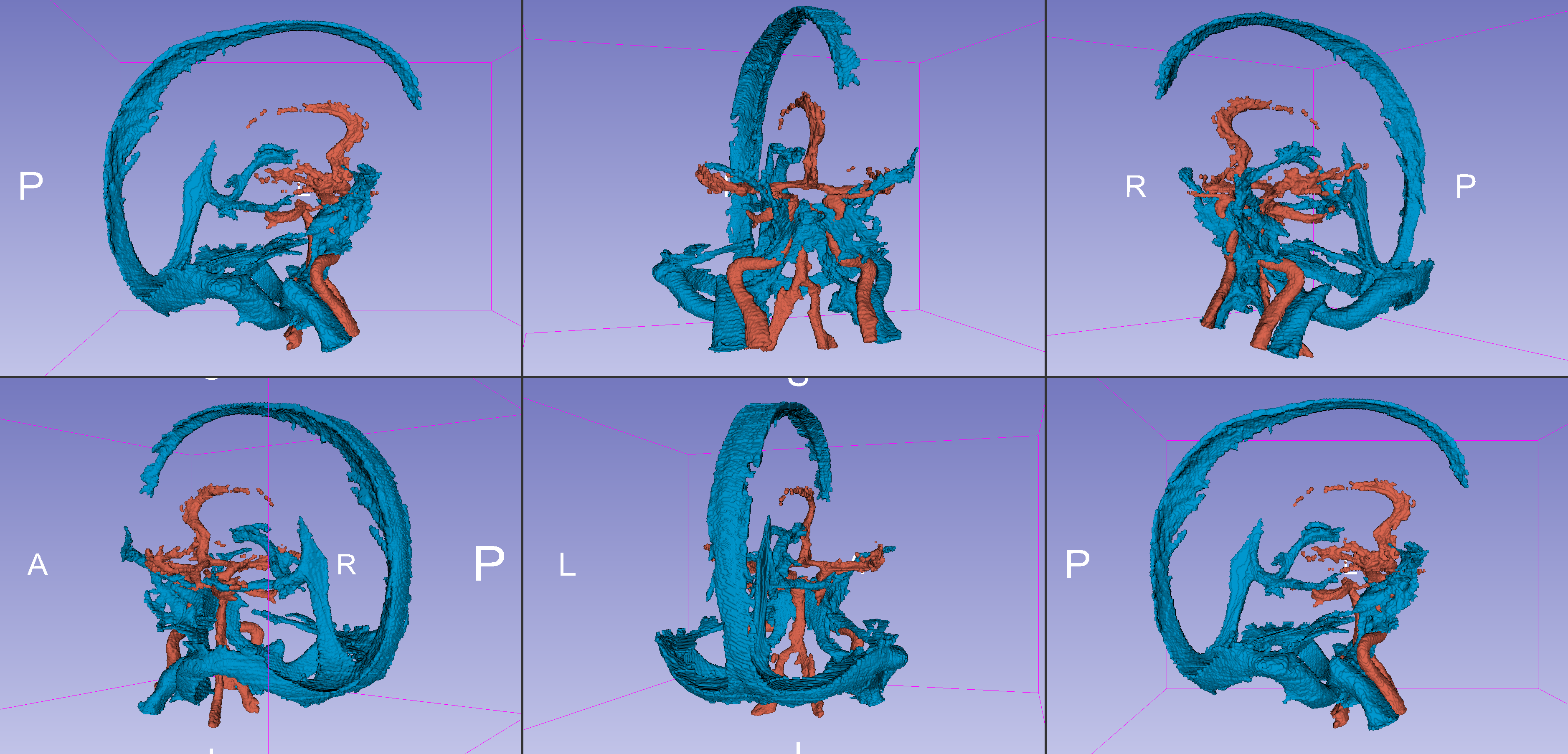}
   \captionv{12}{}
   {Lightbox view of the 3D surface rendering of $\text{ROI}_a^*$ (identified from $T_{as}^*$) in red and $\text{ROI}_v^*$ (identified from $T_{vs}^*$) in blue. 
   \label{fig:reg_seg} 
    }  
    \end{center}
\end{figure}

\begin{figure}[ht]
   \begin{center}
   \includegraphics[width=\textwidth]{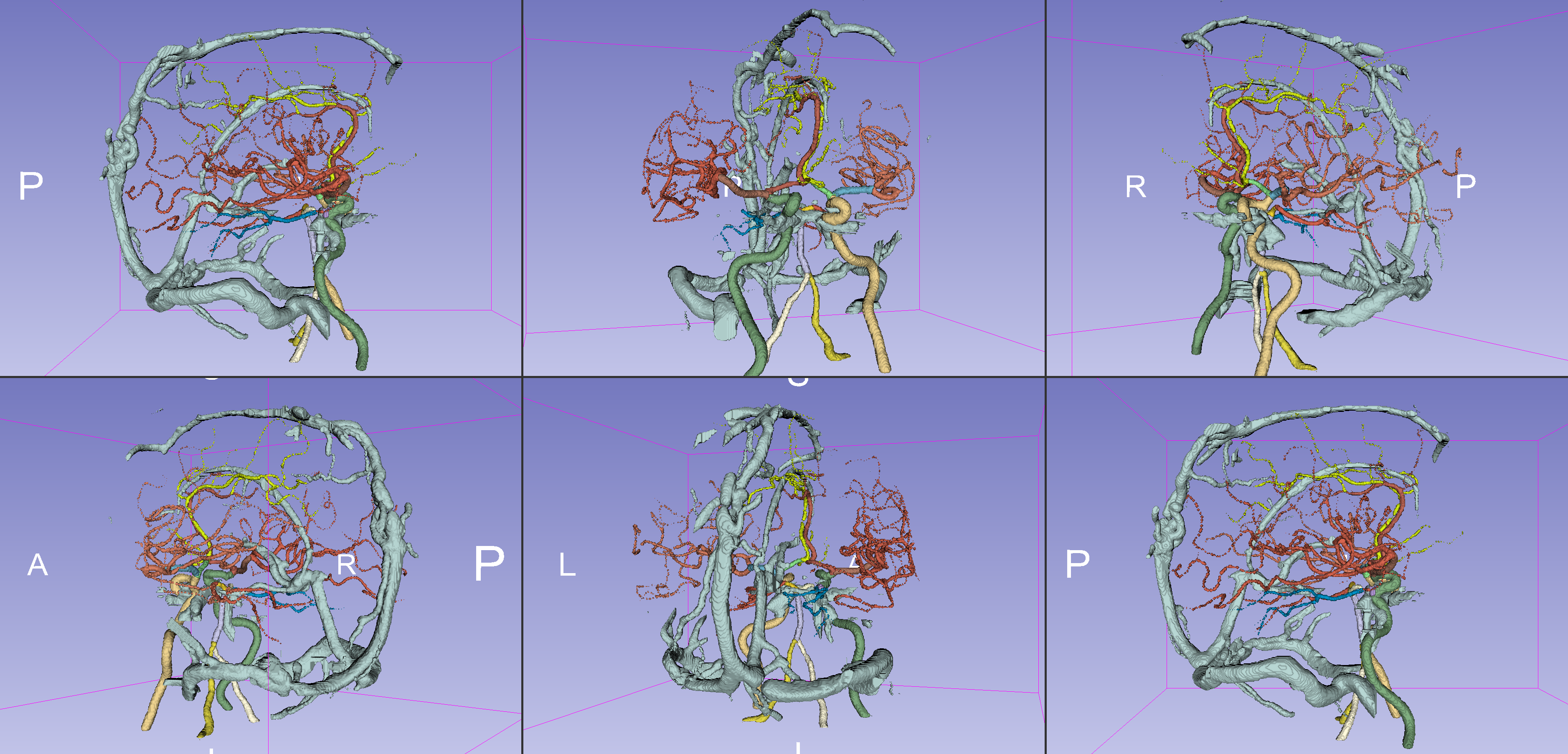}
   \captionv{12}{}
   {Lightbox view of the 3D surface rendering of the ground truth annotations from iCafe. 
   \label{fig:gt} 
    }  
    \end{center}
\end{figure}

    

\begin{figure}[ht]
   \begin{center}
   \includegraphics[clip, trim=3in 0in 1.5in 0in,width=\textwidth]{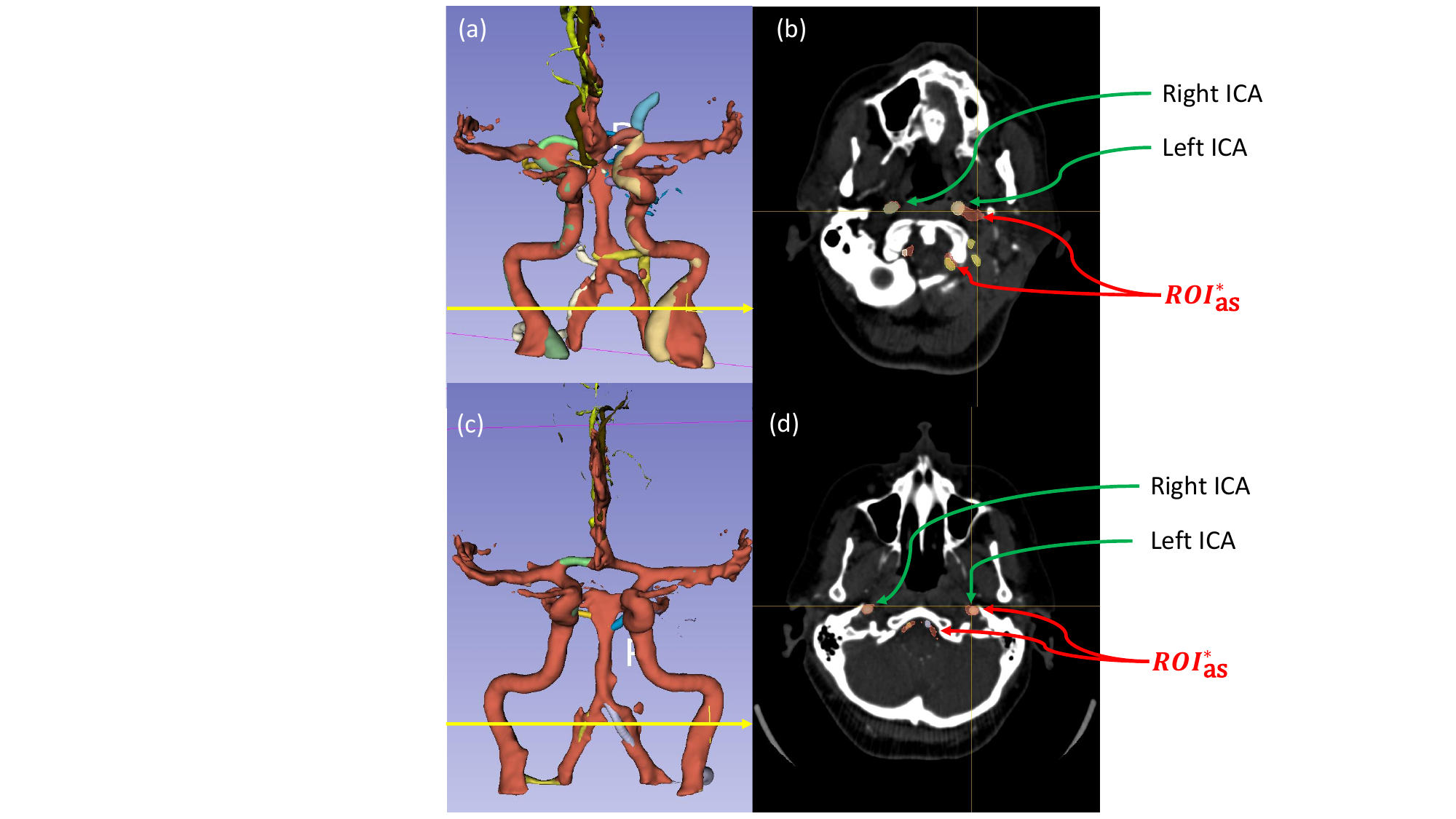}
   \captionv{12}{}
   { Figure shows two examples where registration-based segmentation succeeded and failed, respectively. Panel (a): 3D surface rendering of a poor registration where the ICAs and the vertebral arteries are poorly aligned with the ROI. ICA (R), ICA (L), and M1 (R) were moderately captured with 0.87, 0.7, and 0.72 mDC. VA (L) and VA (R) were poorly captured with 0.33 and 0.21 mDC. Panel (b) 3D surface rendering of a good registration. ICA, BA, A1 (L), M1,  were well segmented (mDC$>$0.97). The VA (L) and VA (R) were moderately segmented with 0.46 and 0.47 mDC, respectively. The yellow line in panels (a) and (c) indicates the location of the axial slice displayed in panels (b) and (d). 
   \label{fig:good_bad_ex} 
    }  
    \end{center}
\end{figure}

\FloatBarrier

\section*{Comparing Registrations to $X_a/X_v$, $S_a/S_v$, and $S_a^*/S_v^*$}

To compare the segmentation metrics for the three sets of registrations, we used the paired $t$-test. In cases where the differences in the mDC metric were not normally distributed, we employed the Wilcoxon signed-rank test. Normality was assessed using the Shapiro-Wilk test with a significance level of $\alpha = 0.05$.

Table~\ref{tab:method1_method2_vs_method3} compares the mDC obtained from registering to $X_a/X_v$(Group 1), $S_a/S_v$ (Group 2), and $S_a^*/S_v^*$ (Group 3). The mDC values for all vessel types significantly improved when using Group 2 compared to Group 1. For instance, the right and left ICAs increased from 0.558 and 0.599 to 0.961 and 0.927 mDC ($p$-value$<$0.001). Bewteen Group 2 and Group 3, the differences were insignificant for most vessels. The change in mDC for a few vessels was close to the significance threshold. These include ICA (L) ($p$-value 0.022), AL (L) ($p$-value 0.011), both left A2+ ($p$-value 0.046), and the right A2+ ($p$-value 0.019), whereas the mDC for veins and CVS had $p$-values close to the significance threshold (0.059 and 0.037, respectively).

\begin{table}[htbp]
\begin{center}
\scriptsize
\captionv{10}{}{Merged Comparison of Vessel Registrations
\label{tab:method1_method2_vs_method3}
\vspace*{2ex}}
\begin{tabular} {|c|c|c|c|c|c|c|c|}
\hline
\multirow{2}{*}{Vessel name} & \multicolumn{3}{c|}{mDC} & \multicolumn{2}{c|}{$p$-values} & \multirow{2}{*}{Samples}\\
\cline{2-6}
& Group 1 & Group 2 & Group 3 & Group 1-Group 2 & Group 2-Group 3 & \\
\hline
ICA (R) & 0.580 & 0.961 & 0.956 & \textbf{$<$0.001}$^{\dagger}$ & 0.065$^{\dagger}$ & 11\\
\hline
ICA (L) & 0.599 & 0.927 & 0.922 & \textbf{$<$0.001}$^{\dagger}$ & \textbf{0.022}$^{\dagger}$ & 11\\
\hline
M1 (R) & 0.404 & 0.861 & 0.862 & \textbf{$<$0.001}$^{\dagger}$ & 0.878$^{\ddagger}$ & 11\\
\hline
M1 (L) & 0.535 & 0.836 & 0.847 & \textbf{0.001}$^{\dagger}$ & 0.878$^{\ddagger}$ & 11\\
\hline
M2+ (R) & 0.065 & 0.126 & 0.130 & \textbf{$<$0.001}$^{\dagger}$ & 0.329$^{\dagger}$ & 11\\
\hline
M2+ (L) & 0.083 & 0.138 & 0.142 & \textbf{0.005}$^{\ddagger}$ & 0.071$^{\dagger}$ & 11\\
\hline
A1 (R) & 0.377 & 0.801 & 0.837 & \textbf{$<$0.001}$^{\dagger}$ & 0.086$^{\ddagger}$ & 11\\
\hline
A1 (L) & 0.527 & 0.860 & 0.861 & \textbf{$<$0.001}$^{\dagger}$ & \textbf{0.011}$^{\ddagger}$ & 11\\
\hline
A2+ (R) & 0.143 & 0.254 & 0.268 & \textbf{0.002}$^{\dagger}$ & \textbf{0.019}$^{\ddagger}$ & 11\\
\hline
A2+ (L) & 0.135 & 0.223 & 0.230 & \textbf{0.003}$^{\dagger}$ & \textbf{0.046}$^{\dagger}$ & 11\\
\hline
AComm & 0.354 & 0.661 & 0.688 & \textbf{0.012}$^{\dagger}$ & 0.113$^{\dagger}$ & 8\\
\hline
VA (R) & 0.248 & 0.498 & 0.500 & \textbf{0.002}$^{\dagger}$ & 0.872$^{\dagger}$ & 11\\
\hline
VA (L) & 0.229 & 0.471 & 0.471 & \textbf{$<$0.001}$^{\ddagger}$ & 0.977$^{\dagger}$ & 11\\
\hline
BA & 0.579 & 0.879 & 0.831 & \textbf{0.001}$^{\dagger}$ & 0.089$^{\ddagger}$ & 11\\
\hline
P1 (R) & 0.628 & 0.816 & 0.816 & \textbf{0.035}$^{\dagger}$ & 0.594$^{\ddagger}$ & 11\\
\hline
P1 (L) & 0.493 & 0.744 & 0.775 & \textbf{0.001}$^{\dagger}$ & 0.092$^{\ddagger}$ & 10\\
\hline
P2+ (R) & 0.163 & 0.318 & 0.324 & \textbf{$<$0.001}$^{\dagger}$ & 0.638$^{\ddagger}$ & 11\\
\hline
P2+ (L) & 0.166 & 0.314 & 0.323 & \textbf{0.001}$^{\dagger}$ & 0.074$^{\ddagger}$ & 11\\
\hline
PComm (R) & 0.060 & 0.426 & 0.408 & \textbf{0.025}$^{\dagger}$ & 0.141$^{\dagger}$ & 6\\
\hline
PComm (L) & 0.210 & 0.517 & 0.543 & \textbf{0.004}$^{\dagger}$ & 0.204$^{\dagger}$ & 6\\
\hline
Vein & 0.548 & 0.801 & 0.805 & \textbf{$<$0.001}$^{\dagger}$ & 0.059$^{\ddagger}$ & 11\\
\hline
CVS & 0.428 & 0.643 & 0.650 & \textbf{$<$0.001}$^{\dagger}$ & \textbf{0.037}$^{\ddagger}$ & 11\\
\hline
\end{tabular}
\end{center}
\vspace*{2ex}
\footnotesize{$^{\dagger}$ Paired t-test; $^{\ddagger}$ Wilcoxon rank test}
\end{table}

\newpage

\begin{figure}
    \centering
    \includegraphics[width = \textwidth]{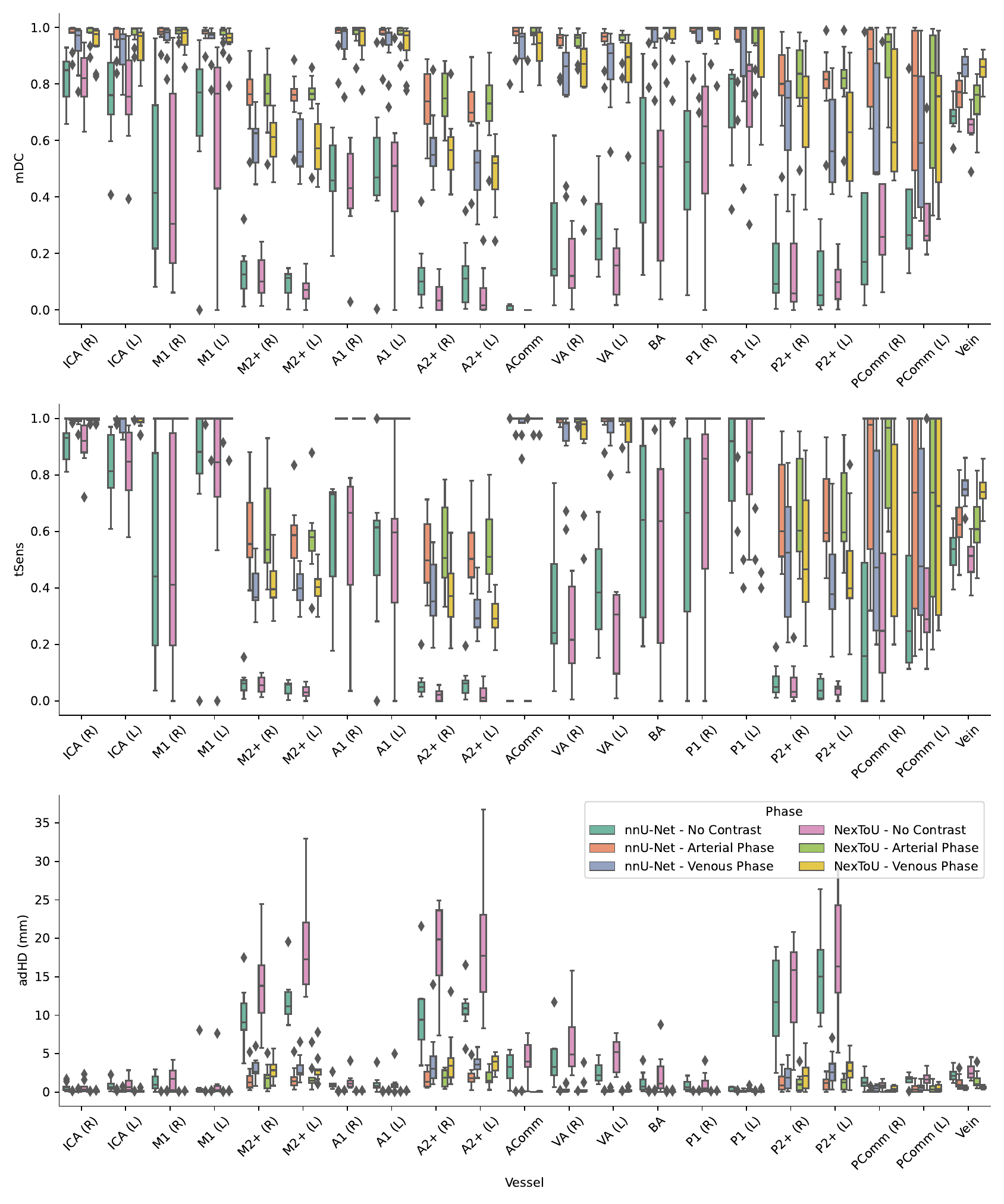}
    \caption{Modified Dice, Topology sensitivity, and average directed Hausdorff distance for all vessel labels.}
    \label{fig:dice_coeff}
\end{figure}
\newpage
\begin{figure}[ht]

    \centering
    \includegraphics[clip, trim=1.4in 0in 0.5in 0in, width=\textwidth, height=0.9\textheight, keepaspectratio]{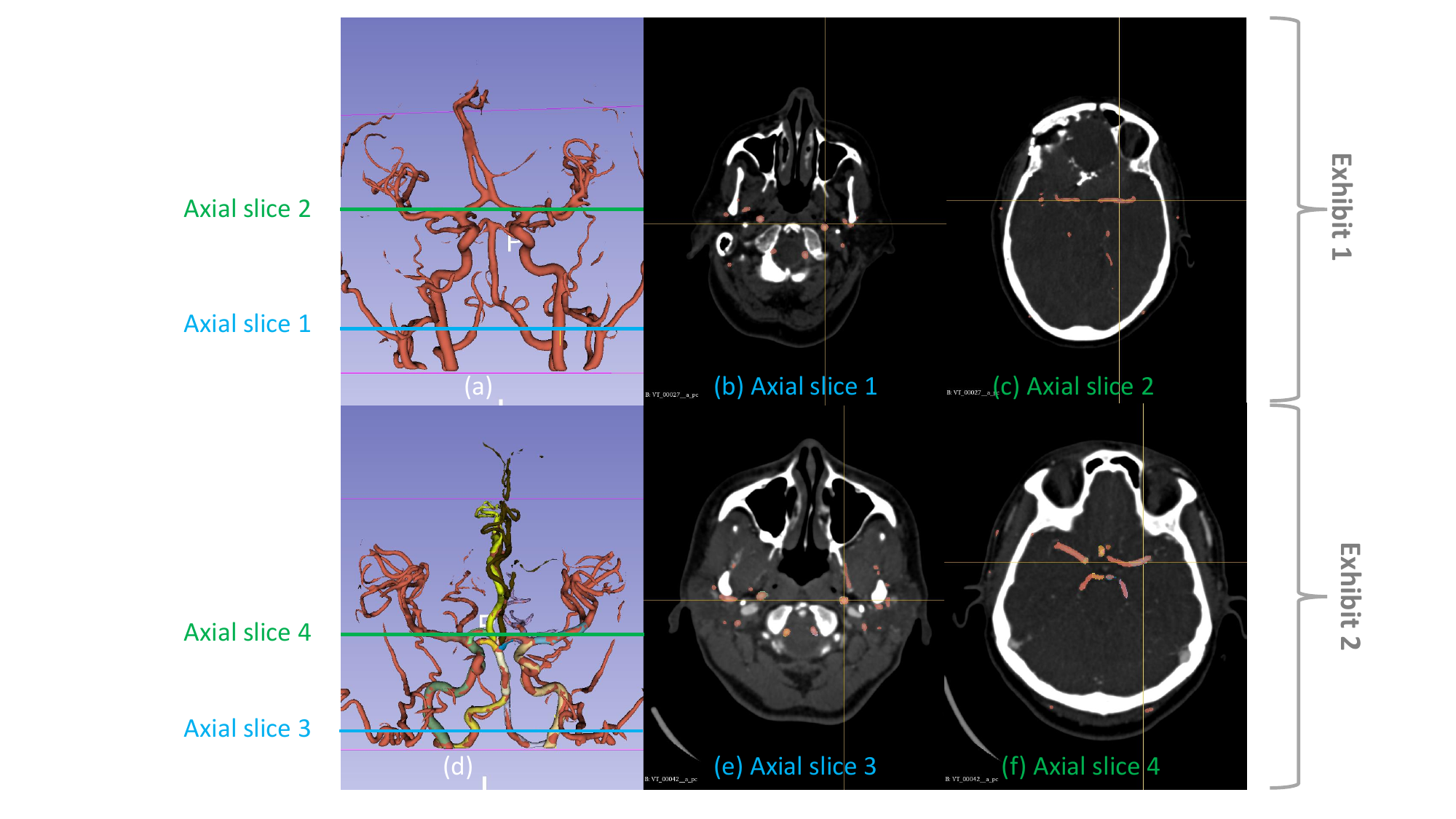}
    \caption{}{Two exhibits with the highest and the lowest average mDC of 0.95 and 0.73.\label{fig:good_bad_examples}}
\end{figure}